\documentclass[twocolumn,showpacs,prb,aps,amsmath,amssymb,superscriptaddress]{revtex4}
\usepackage[dvipdfmx]{graphicx}
\usepackage{bm}
\usepackage[dvipdfmx]{color}

\usepackage{color}

\begin{document}

\title{
Dimer correlation amplitudes and dimer excitation gap in spin-1/2 XXZ
and Heisenberg chains
}
\author{Toshiya Hikihara}
\affiliation{Faculty of Science and Technology, Gunma University, 
Kiryu, Gunma 376-8515, Japan}
\author{Akira Furusaki}
\affiliation{Condensed Matter Theory Laboratory, RIKEN,
Wako, Saitama 351-0198, Japan}
\affiliation{RIKEN Center for Emergent Matter Science 
(CEMS), Wako, Saitama, 351-0198, Japan}
\author{Sergei Lukyanov}
\affiliation{NHETC, Department of Physics and Astronomy, 
Rutgers University, Piscataway, NJ 08855-0849, USA}
\date{\today}

\begin{abstract}
Correlation functions of dimer operators,
the product operators of spins on two adjacent sites,
are studied in the spin-$\frac{1}{2}$ XXZ chain in the critical regime.
The amplitudes of the leading oscillating terms in the dimer correlation
functions are determined with high accuracy
as functions of the exchange anisotropy
parameter and the external magnetic field,
through the combined use of bosonization and
density-matrix renormalization group methods.
In particular,
for the antiferromagnetic Heisenberg model with SU(2) symmetry,
logarithmic corrections to the dimer correlations
due to the marginally-irrelevant operator are studied, and
the asymptotic form of the dimer correlation function is obtained.
The asymptotic form of the spin-Peierls excitation gap
including logarithmic corrections is also derived.
\end{abstract}

\maketitle

\section{Introduction}
The one-dimensional (1D) model of $S=\frac12$ spins
with anisotropic exchange interaction, the spin-$\frac12$ XXZ chain,
is a basic model in quantum magnetism.
Its Hamiltonian is given by
\begin{equation}
\mathcal{H} = J \sum_{l=1}^{L-1}
( S^x_l S^x_{l+1} + S^y_l S^y_{l+1} + \Delta S^z_l S^z_{l+1})
- h \sum_{l=1}^{L} S^z_l,
\label{eq:HamXXZ}
\end{equation}
where ${\bm S}_l = \left( S^x_l, S^y_l, S^z_l \right)$ is
the spin-$\frac12$ operator at the $l$th site, $L$ is the number of spins,
$\Delta$ is the anisotropy parameter, and $h$ is the external magnetic field.
The exchange-coupling constant is assumed to be positive,
$J > 0$.
The XXZ chain is an important toy model,
from both experimental and theoretical viewpoints,
for understanding magnetic properties of various (quasi-)1D materials.

An intriguing feature of the XXZ chain is that
it realizes a quantum-critical Tomonaga-Luttinger liquid (TLL)
for a large region in the two-dimensional parameter space
$(\Delta,h)$.\cite{Giamarchi-text,GogolinNT-text,Affleck-text}
In the TLL phase of the XXZ chain, strong quantum fluctuations
prevent spontaneous breaking of
continuous symmetries even at zero temperature;
in the resulting critical ground state, correlation functions
have power-law dependence on the distance or time.
For example, the equal-time spin-spin correlation functions
in the ground state of the XXZ chain in the TLL phase
have the asymptotic forms,\cite{LutherP1975}
\begin{subequations}
\label{eq:SScor}
\begin{align}
&\langle S^x_l S^x_{l+r} \rangle =
A^x_0 \frac{(-1)^r}{r^\eta}
- A^x_1 \frac{\cos(2\pi M r)}{r^{\eta+1/\eta}} + \cdots,
\label{eq:S+S-cor} \\
&\langle S^z_l S^z_{l+r} \rangle - M^2 =
A^z_1 \frac{(-1)^r \cos(2\pi M r)}{r^{1/\eta}} -\frac{1}{4\pi^2\eta r^2} 
+ \cdots
\nonumber \\
\label{eq:SzSzcor}
\end{align}
\end{subequations}
for long distance $r$ in the bulk ($1\ll r \ll L$, $l \approx L/2$),
where $M=\langle S^z_l\rangle$ is the magnetization per spin
and $\langle \cdots \rangle$ denotes the expectation value in the ground state.
The parameter $\eta$ in the exponents can be obtained exactly by solving
integral equations from the Bethe ansatz, and its explicit solution
at $M=0$ (i.e., $h=0$) is given by\cite{LutherP1975}
\begin{eqnarray}
\eta = 1 - \frac{1}{\pi} \arccos(\Delta).
\label{eq:eta_zeroM}
\end{eqnarray}
The amplitudes $A^x_0$, $A^x_1$, and $A^z_1$ have been determined
as functions of $\Delta$ and $M$.\cite{LukyanovZ1997,Lukyanov1998,Lukyanov1999,LukyanovT2003,ShashiPCI2012,HikiharaF1998,HikiharaF2001,HikiharaF2004}
The dynamical spin-structure factors of the XXZ chain have also been
calculated.\cite{CauxKSW2012}

In this paper, we focus our attention on correlation functions
of the product of two adjacent spins,
\begin{subequations}
\label{eq:Odim}
\begin{align}
\mathcal{O}_{\rm d}^{\pm}(l) &=
\frac{1}{2} (S^x_l S^x_{l+1} + S^y_l S^y_{l+1}) 
\nonumber \\
&= \frac{1}{4} (S^+_l S^-_{l+1} + S^-_l S^+_{l+1}),
\label{eq:Odim+-} \\
\mathcal{O}_{\rm d}^{z}(l) &= S^z_l S^z_{l+1},
\label{eq:Odimzz}
\end{align}
\end{subequations}
where $S^\pm_l = S^x_l \pm i S^y_l$.
We call them dimer operators
[their superposition
$2\mathcal{O}_{\rm d}^{\pm}(l)+\Delta \mathcal{O}_{\rm d}^{z}(l)$
is the ``energy operator''].
One can show, using the bosonization method,
that the correlation functions of the dimer operators
in the critical TLL phase of the XXZ chain have
the asymptotic forms\cite{Giamarchi-text,GogolinNT-text,Affleck-text}
\begin{align}
&\langle\mathcal{O}_{\rm d}^a(l)\mathcal{O}_{\rm d}^a(l+r)\rangle
-\langle\mathcal{O}_{\rm d}^a(l)\rangle
 \langle\mathcal{O}_{\rm d}^a(l+r)\rangle
\nonumber \\
&~~~= B^a_1 \frac{(-1)^r \cos(2\pi M r)}{r^{1/\eta}}
+ \frac{B_2^a}{r^4}
+ B_3^a\frac{\cos(4\pi Mr)}{r^{4/\eta}} + \cdots,
\label{eq:Odim-cor}
\end{align}
where $a = \pm, z$.
The exponent $1/\eta$ of the first term on the right-hand side is
the same as that of the first term in Eq.\ (\ref{eq:SzSzcor}).
Thus, the oscillating term in the dimer correlation function is as
important as the oscillating component in the longitudinal spin correlation
in the TLL phase.
These two terms are related to the same vertex operators
$\exp(\pm i\sqrt{2\pi}\phi)$ in the low-energy effective theory
[see related discussion below Eq.\ (\ref{eq:O_boson_formula})].
The dimer correlation is also important as a measure of
the instability towards spin-Peierls order.\cite{CrossF1979}
In the spin-Peierls phase where there is a small alternation in the magnitude
of the exchange interaction $J$, spin excitations have an energy gap
whose size and scaling are directly related to the dimer correlation
in the spin chain without the alternation in $J$
[the first term in Eq.\ (\ref{eq:Odim-cor})].
To the best of our knowledge,
the exact values of the correlation amplitudes $B_1^a$ are not known,
and so far they are only numerically estimated from
the exact diagonalization of small systems.\cite{TakayoshiS2010}
Experimentally, the dynamical structure factor of the dimer operators
can be probed in the optical absorption spectrum\cite{Suzuura1996} and
resonant inelastic x-ray scattering.\cite{Ament2011,NagaoI2007}
Accurate evaluation of the dynamical structure factor of
the dimer operators has been performed using the algebraic Bethe ansatz
combined with numerical computation.\cite{KlauserMCB2011,KlauserMC2012}

The purpose of this paper is to numerically determine the amplitudes $B^a_1$
of the leading term in Eq.\ (\ref{eq:Odim-cor})
to very high accuracy.
This is achieved by combining the powerful analytical and numerical approaches
available for 1D systems:
bosonization and density-matrix renormalization group (DMRG) methods.
The bosonization method provides the low-energy effective theory
of the XXZ spin chain.\cite{Giamarchi-text,GogolinNT-text,Affleck-text}
We calculate the ground-state expectation values of the dimer operators
in finite spin chains with
open boundaries 
using the bosonization and DMRG methods.
The numerical data from the DMRG calculation
are fitted to the corresponding formulas from the bosonization;
this allows us to obtain accurate numerical
estimates of the amplitudes $B_1^a$.

Another important result of this work concerns the dimer correlations in
the SU(2) symmetric case where $\Delta=1$ and $h=0$ in Eq.\ (\ref{eq:HamXXZ}).
In this case, a marginally-irrelevant operator in the low-energy effective
theory leads to logarithmic corrections in various physical quantities.
An interesting example is a spin excitation gap in the antiferromagnetic
Heisenberg spin chain with weak bond alternation.
Since the gap is directly related to the dimer correlation,
we can determine, from the scaling analysis of the excitation gap,
the amplitude of the leading dimer correlation
with a multiplicative logarithmic correction;
our result is consistent with
a recent numerical estimate reported in Ref.~\onlinecite{VekuaS2016}.
We also derive the asymptotic form of the excitation gap in
the bond-alternating Heisenberg chain.

The organization of the rest of the paper is as follows.
In Sec.\ \ref{sec:XXZ_zeroM} we focus on the case of vanishing
magnetization $M=0$ ($h=0$) and easy-plane anisotropy $|\Delta| < 1$.
The correlation amplitudes $B^a_1$ are obtained as a function of
the anisotropy $\Delta$.
In Sec.\ \ref{sec:SU2} we discuss the SU(2) symmetric case and
derive the asymptotic forms of the dimer correlation function
and the spin-Peierls excitation gap 
with the logarithmic correction.
In Sec.\ \ref{sec:finiteM} we present the correlation amplitudes $B^a_1$
in the partially-polarized case $0 < M < 1/2$.
Section \ref{sec:conc} is devoted to a summary and concluding remarks.

\section{XXZ chain in zero magnetic field}\label{sec:XXZ_zeroM}
\subsection{Theory}\label{subsec:XXZ_theory}

In this section, we consider the XXZ model in Eq.\ (\ref{eq:HamXXZ})
for $-1 < \Delta < 1$ and $h=0$.
In this parameter regime, the low-energy effective theory is
a free-boson theory, i.e., the Gaussian model,
\begin{equation}
\widetilde{\mathcal{H}}_0=\frac{v}{2}\int^{L+1}_0 dx
\left[
\frac{1}{\eta}:\!\left(\frac{d\theta}{dx}\right)^2\!:
+
\eta:\!\left(\frac{d\phi}{dx}\right)^2\!:
\right],
\label{eq:Gaussian}
\end{equation}
where $\phi(x)$ and $\theta(x)$ are bosonic fields that are dual to each other
and satisfy the commutation relation $[\phi(x),d\theta(y)/dy]=i\delta(x-y)$.
The field $\phi(x)$ is compactified as $\phi+\sqrt{2\pi} \equiv \phi$.
The operators in the integrand in Eq.\ (\ref{eq:Gaussian})
are normal-ordered, as indicated by the colons. 
The parameter $\eta$ is given by Eq.\ (\ref{eq:eta_zeroM}), and
the renormalized spin velocity $v$ is related to $\Delta$ (and $\eta$)
as\cite{dCloizeauxP1962,BogoliubovIK1986}
\begin{eqnarray}
v = \frac{\pi\sqrt{1-\Delta^2}}{2\arccos(\Delta)}J
=\frac{\sin(\pi\eta)}{2(1-\eta)}J.
\label{eq:v_zero_field}
\end{eqnarray}
We set the lattice spacing to unity
so that the continuous coordinate $x$ can be identified with
the lattice index $l$.
We note that in the effective Hamiltonian (\ref{eq:Gaussian}),
we have discarded symmetry-allowed operators which are irrelevant
in the renormalization-group sense.
Among those operators, the leading irrelevant term
$g \cos(\sqrt{8\pi}\phi)$
has scaling dimension $2/\eta$ and becomes
marginally irrelevant at the SU(2)-symmetric point ($\Delta=1$ and $h=0$),
yielding the logarithmic corrections.
Therefore, our results presented below
(in this section and Sec.\ \ref{sec:finiteM})
may include systematic errors near the SU(2) point
due to the leading irrelevant cosine term.
The SU(2)-symmetric case will be discussed in Sec.\ \ref{sec:SU2},
where the effect of the marginally irrelevant perturbation
$g\cos(\sqrt{8\pi}\phi)$ is taken into account.

The dimer operators defined in Eq.\ (\ref{eq:Odim}) are expressed
in terms of the bosonic fields as\cite{Giamarchi-text,GogolinNT-text,Affleck-text}
\begin{align}
\mathcal{O}_{\rm d}^a(l)
=&\, c_0^a + c_1^a(-1)^l\cos[\sqrt{2\pi}\phi(x_l)]
\nonumber\\
& + c_\phi^a :\!\! \left(\frac{d\phi(x_l)}{dx}\right)^2 \!:
+\, c_\theta^a :\!\! \left(\frac{d\theta(x_l)}{dx}\right)^2 \!:
\nonumber\\
& + c_g^a \cos[\sqrt{8\pi}\phi(x_l)]
+\cdots,
\label{eq:O_boson_formula}
\end{align}
where $x_l=l+\frac{1}{2}$ is the center position of
two spins forming dimer operators.
Note that the second term on the right-hand side is a cosine of
the field $\phi$, so that the ground-state expectation value of
Eq.\ (\ref{eq:O_boson_formula})
with the Dirichlet boundary condition (\ref{eq:Dirichlet}) correctly yields
the Friedel oscillations near the open boundaries,
as we will see in
Eqs.\ (\ref{eq:Odim-finite}) and (\ref{eq:Friedel}).
Incidentally, the bosonization of the $z$-component of the spin operator,
$S^z_l$, has $(-1)^l\sin(\sqrt{2\pi}\phi)$.\cite{HikiharaF1998,HikiharaF2004}
A higher-order term ($\propto\cos\sqrt{8\pi}\phi$) is also included
in Eq.\ (\ref{eq:O_boson_formula}) for later convenience.
Our task is to determine the coefficients
in Eq.\ (\ref{eq:O_boson_formula}).
Among them, those of the uniform terms
($c_0^a$, $c_\phi^a$, $c_\theta^a$, and $c_g^a$)
can be obtained exactly as follows.

Since a linear combination of the dimer operators,
$2\mathcal{O}_{\rm d}^{\pm}+\Delta \mathcal{O}_{\rm d}^{z}$,
is nothing but the exchange interaction in the XXZ model (\ref{eq:HamXXZ})
at $h=0$, the coefficients of the uniform terms
in Eq.\ (\ref{eq:O_boson_formula}) are related to the ground-state energy
and the parameters in the low-energy effective Hamiltonian
of the model.
Then, using the Hellmann-Feynman theorem, the coefficients $c_0^a$
are related to the ground-state energy density $e_0$ 
of the XXZ chain,
\begin{equation}
c_0^z = \frac{1}{J}\frac{\partial e_0}{\partial\Delta},
\qquad
c_0^\pm = \frac{1}{2J}\!
\left(e_0 - \Delta \frac{\partial e_0}{\partial\Delta}\right).
\label{eq:c0_to_e0}
\end{equation}
Substituting the exact ground-state energy density
$e_0$ obtained from the Bethe ansatz,\cite{YangY1966A,YangY1966B,Baxter1972} 
\begin{equation}
\frac{e_0}{J}=
-\frac{\sin(\pi\eta)}{\pi}
\int^\infty_0\!\!
\frac{\sinh(\eta t)dt}{\sinh(t)\cosh[(1-\eta)t]} 
- \frac{\cos(\pi \eta)}{4},
\label{eq:e0_zero_field}
\end{equation}
into Eq. (\ref{eq:c0_to_e0}) gives
\begin{subequations}
\label{eq:c0_zero_field}
\begin{align}
c_0^z &=
\frac{1}{4} - \frac{\cos(\pi\eta)}{\pi\sin(\pi\eta)}I_1
- \frac{1}{\pi^2}I_2,
\label{eq:c0zz_zero_field}
\\
c_0^\pm &=
-\frac{1}{2\pi\sin(\pi\eta)}I_1 -\frac{\cos(\pi\eta)}{2\pi^2}I_2,
\label{eq:c0+-_zero_field}
\end{align}
\end{subequations}
where the integrals $I_1$ and $I_2$ are given by
\begin{subequations}
\begin{align}
I_1 &=
\int_0^\infty \!
\frac{\sinh(\eta t) dt}{\sinh(t)\cosh[(1-\eta)t]},
\\
I_2 &=
\int_0^\infty \!
\frac{t\cosh(t) dt}{\sinh(t)\cosh^2[(1-\eta)t]}.
\end{align}
\end{subequations}
Similarly, comparing the third and fourth terms
in Eq.\ (\ref{eq:O_boson_formula}) with the Hamiltonian density of
the Gaussian model (\ref{eq:Gaussian}),
one finds that the coefficients $c_\phi^a$ and $c_\theta^a$ are expressed
in terms of the spin velocity $v$ and the parameter $\eta$ as
\begin{subequations}
\label{eq:cphi_and_ctheta_to_e0}
\begin{align}
c_\phi^z &=
\frac{1}{2J}\frac{\partial v\eta}{\partial\Delta},
\quad
c_\phi^\pm=\frac{1}{4J}\!\left(
v\eta-\Delta\frac{\partial v\eta}{\partial\Delta}
\right),
\\
c_\theta^z &=
\frac{1}{2J}\frac{\partial(v/\eta)}{\partial\Delta},
\quad
c_\theta^\pm=\frac{1}{4J}\!\left(
\frac{v}{\eta}-\Delta\frac{\partial(v/\eta)}{\partial\Delta}
\right).
\end{align}
\end{subequations}
These relations, together with Eqs.\ (\ref{eq:eta_zeroM}) and
(\ref{eq:v_zero_field}), determine $c_\phi^a$ and $c_\theta^a$:
\begin{subequations}
\begin{align}
c_\phi^z=&
\frac{\pi\eta(1-\eta)\cos(\pi\eta)+\sin(\pi\eta)}
{4\pi(1-\eta)^2\sin(\pi\eta)},
\label{c_phi^z}
\\
c_\phi^\pm=&
\frac{2\pi\eta(1-\eta)+\sin(2\pi\eta)}
{16\pi(1-\eta)^2\sin(\pi\eta)},
\label{c_phi^pm}
\\
c_\theta^z=&
\frac{\pi\eta(1-\eta)\cos(\pi\eta)+(2\eta-1)\sin(\pi\eta)}
{4\pi\eta^2(1-\eta)^2\sin(\pi\eta)},
\label{c_theta^z}
\\
c_\theta^\pm=&
\frac{2\pi\eta(1-\eta)+(2\eta-1)\sin(2\pi\eta)}
{16\pi\eta^2(1-\eta)^2\sin(\pi\eta)}.
\label{c_theta^pm}
\end{align}
\end{subequations}
Note that these coefficients diverge at the SU(2) isotropic limit $\eta \to 1$
as $c_\phi^a, c_\theta^a \propto (\eta-1)^{-1}$, which signals the appearance of logarithmic
corrections [$\propto(\ln r)^2$] in the uniform term ($\propto1/r^4$)
of the dimer correlation function in Eq.\ (\ref{eq:Odim-cor})
(see also Ref.~\onlinecite{VekuaS2016}).
Incidentally, $c_g^a$ are related to the coupling constant $g$
of the irrelevant perturbation $g\cos(\sqrt{8\pi}\phi)$
to the Gaussian Hamiltonian,
\begin{equation}
c_g^z=\frac{1}{J}\frac{\partial g}{\partial\Delta},
\quad
c_g^\pm=\frac{1}{2J}\!\left(g-\Delta\frac{\partial g}{\partial\Delta}\right).
\label{eq:cg_to_e0}
\end{equation}
The explicit form of $g$ in the effective Hamiltonian for
$-1<\Delta<1$ and $h=0$ is given in Ref.~\onlinecite{LukyanovT2003}
and used in numerical studies.\cite{FurusakiH,Furukawa}
We will not consider the higher-order harmonics
$c_g^a\cos(\sqrt{8\pi}\phi)$ anymore in this section,
because its contribution ($\propto r^{-4/\eta}$) in Eq.\ (\ref{eq:Odim-cor})
decays faster than the other terms for $\eta<1$.

In contrast to the coefficients of the uniform part discussed above,
the exact formula for the coefficients $c_1^a$
in Eq.\ (\ref{eq:O_boson_formula})
is not available,
except for the free-fermion point $\Delta=0$,
\begin{equation}
c_1^\pm(\Delta=0) = \frac{1}{2\pi},
\qquad
c_1^z(\Delta=0) = \frac{2}{\pi^2}.
\end{equation}
In order to evaluate $c_1^a$, we consider
Friedel oscillations in the expectation values of the dimer operators $\mathcal{O}^a_\mathrm{d}(l)$
near the open boundaries,
which can be easily studied by applying the DMRG method
to finite open chains.
We also calculate the ground-state expectation values of
the dimer operators
using the bosonization method.
In the effective theory, the presence of open boundaries can be
taken into account by imposing the Dirichlet boundary conditions
on the bosonic field $\phi(x)$,\cite{HikiharaF1998,HikiharaF2001,HikiharaF2004,EggertA1992}
\begin{equation}
\phi(0)=\phi(L+1)=0.
\label{eq:Dirichlet}
\end{equation}
Since the low-energy theory is the Gaussian model in Eq.\ (\ref{eq:Gaussian}),
we expand the bosonic fields with harmonic oscillator modes as
\begin{subequations}
\label{eq:mode_expansion}
\begin{align}
\sqrt{\eta}\phi(x) &=
\frac{x}{L+1}\phi_0
+\sum_{n=1}^\infty e^{-\alpha n/2}\frac{\sin q_nx}{\sqrt{\pi n}}
  \left(a_n^{}+a_n^\dagger\right),
\label{eq:phi_mode}
\\
\frac{1}{\sqrt{\eta}}\theta(x) &=
\theta_0
+i\sum_{n=1}^\infty e^{-\alpha n/2}\frac{\cos q_nx}{\sqrt{\pi n}}
  \left(a_n^{}-a_n^\dagger\right),
\label{eq:theta_mode}
\end{align}
\end{subequations}
where $q_n=\pi n/(L+1)$, $[\theta_0,\phi_0]=i$,
and $[a_m^{},a_n^\dagger]=\delta_{m,n}$.
The parameter $\alpha$ is a small positive constant
that is introduced for regularization.
The fields $\phi(x)$ and $\theta(x)$ in Eq.\ (\ref{eq:mode_expansion})
satisfy the commutation relation
$[ \phi(x), \theta(y)] = -(i/2)[1+{\rm sgn}(x-y)]$.
The ground state $|0\rangle$ is a vacuum of the bosons $a_n$
and the zero mode $\phi_0$: $a_n |0 \rangle = \phi_0 |0 \rangle = 0$.

Using the mode expansions in Eq.\ (\ref{eq:mode_expansion}),
the ground-state expectation values of the operators that appear in
Eq.\ (\ref{eq:O_boson_formula}) can be obtained as
\begin{subequations}
\label{eq:expectation_gs}
\begin{align}
&\langle\cos[\sqrt{2\pi}\phi(x)]\rangle =
\frac{1}{[f(2x)]^{1/2\eta}},
\label{eq:cosphi_gs} \\
&\eta
\langle[d\phi(x)/dx]^2\rangle =
- \frac{\pi}{24(L+1)^2} - \frac{1}{2\pi[f(2x)]^2},
\label{eq:dphi2_gs} \\
&\frac{1}{\eta}
\langle[d\theta(x)/dx]^2\rangle =
- \frac{\pi}{24(L+1)^2} + \frac{1}{2\pi[f(2x)]^2}.
\label{eq:dtheta2_gs}
\end{align}
\end{subequations}
Here we have defined
\begin{equation}
f(x) = \frac{2(L+1)}{\pi} \sin\!\left(\frac{\pi |x|}{2(L+1)}\right),
\label{eq:f}
\end{equation}
which is simplified to $f(x)=|x|$ in the thermodynamic limit $L\to\infty$.
We have used the regularization
\begin{equation}
\sum_{n=1}^\infty\frac{e^{-\alpha n}}{n}(1-\cos q_nx)=\ln[f(x)]
\end{equation}
in Eq.\ (\ref{eq:cosphi_gs}), such that
the two-point function of vertex operators has the form
\begin{equation}
\langle e^{i\sqrt{2\pi}\mu\phi(x)}e^{-i\sqrt{2\pi}\mu\phi(y)}\rangle
=|x-y|^{-\mu^2/\eta}
\end{equation}
in the bulk limit,
$1 \ll |x-y| \ll L$, $x \approx L/2$, $y \approx L/2$.
Note that we have not normal-ordered
the operators on the left-hand side
of Eqs.\ (\ref{eq:dphi2_gs}) and (\ref{eq:dtheta2_gs}),
so that we can obtain
the finite-size corrections $\propto1/(L+1)^2$
coming from the zero-point energy of the harmonic oscillators.
In this calculation
we have used $\zeta(-1)=-1/12$ and taken the $\alpha\to0$ limit,
assuming that singular contributions proportional to $\alpha^{-2}$
are already included in the ground-state energy density $e_0$.

From Eqs.\ (\ref{eq:O_boson_formula}) and (\ref{eq:expectation_gs}),
we find that the ground-state expectation values of the dimer operators
in finite open chains are given by
\begin{align}
\langle \mathcal{O}_{\rm d}^a (l) \rangle =&\,
c_0^a + \frac{(-1)^l c_1^a}{[f(2l+1)]^{1/2\eta}} 
\nonumber \\
&- \frac{\pi^2 c_2^a}{12(L+1)^2} - \frac{\bar{c}_2^a}{[f(2l+1)]^2} + \cdots.
\label{eq:Odim-finite}
\end{align}
The constants $c_0^a$ are given in Eq.\ (\ref{eq:c0_zero_field}).
We note that $c_1^a$ is positive in the open spin chains (\ref{eq:HamXXZ}).
The coefficients $c_2^a$ and $\bar{c}_2^a$ are related to
$c_\phi$ and $c_\theta$ by
\begin{equation}
c_2^a=\frac{1}{2\pi}\!\left(\frac{c_\phi^a}{\eta}+\eta c_\theta^a\right),
\quad
\bar{c}_2^a=\frac{1}{2\pi}\!\left(\frac{c_\phi^a}{\eta}-\eta c_\theta^a\right),
\end{equation}
and are written explicitly as
\begin{subequations}
\label{eq:c2_and_barc2_zero_field}
\begin{align}
c_2^\pm &=
\frac{\sin(2\pi\eta)+2\pi(1-\eta)}{16\pi^2(1-\eta)^2\sin(\pi\eta)},
\label{eq:c2+-} \\
c_2^z &=
\frac{\sin(\pi\eta)+\pi(1-\eta)\cos(\pi\eta)}{4\pi^2(1-\eta)^2\sin(\pi\eta)},
\label{eq:c2zz} \\
\bar{c}_2^\pm &=
\frac{\cos(\pi\eta)}{8\pi^2\eta(1-\eta)},
\label{eq:barc2+-} \\
\bar{c}_2^z &=
\frac{1}{4\pi^2\eta(1-\eta)}.
\label{eq:barc2zz}
\end{align}
\end{subequations}
We will use these results in the next section to estimate the unknown
coefficients $c_1^a$ from numerical data.

We note that Eq.\ (\ref{eq:Odim-finite}) is simplified to
\begin{equation}
\langle\mathcal{O}^a_\mathrm{d}(l)\rangle
= c_0^a + \frac{(-1)^lc_1^a}{(2l)^{1/2\eta}}-\frac{\bar{c}^a_2}{(2l)^2}
+\cdots
\label{eq:Friedel}
\end{equation}
for $1 \ll l \ll L$.
This should be contrasted with the two-point functions of
the dimer operators in Eq.\ (\ref{eq:Odim-cor}),
which are calculated in the bulk (away from boundaries).
The boundary exponents in Eq.\ (\ref{eq:Friedel}) are half
the bulk exponents in Eq.\ (\ref{eq:Odim-cor}).

Finally, the asymptotic forms of the dimer correlation functions
[Eq.\ (\ref{eq:Odim-cor})] can be derived by calculating 
the correlation functions in finite open chains
using Eqs.\ (\ref{eq:O_boson_formula}) and (\ref{eq:mode_expansion})
and taking the thermodynamic limit $L \to \infty$.
The correlation amplitudes in Eq.\ (\ref{eq:Odim-cor}) are given
in terms of the coefficients $c_j^a$ by
\begin{align}
B_1^a&=\frac{(c_1^a)^2}{2},
\qquad
B_2^a=\frac{1}{2\pi^2}\!
\left[\left(\frac{c_\phi^a}{\eta}\right)^2+(\eta c_\theta^a)^2\right],
\nonumber\\
B_3^a&=\frac{(c_g^a)^2}{2}.
\end{align}

\subsection{Numerical results}\label{subsec:numerics_zeroM}
In this section, we present numerical results on the ground-state expectation
values of the
dimer operators in the XXZ chain (\ref{eq:HamXXZ}) with open boundaries
at zero magnetic field $h=0$.
The numerical data shown here and in the following sections
were obtained using the DMRG method.
The number of block states required to achieve a desired accuracy
depends on the model parameters.
We typically kept a few hundred states ($555$ states in the most severe case) 
and checked that the obtained data had enough accuracy for the
subsequent analysis described below.

\begin{figure}
\begin{center}
\includegraphics[width=70mm]{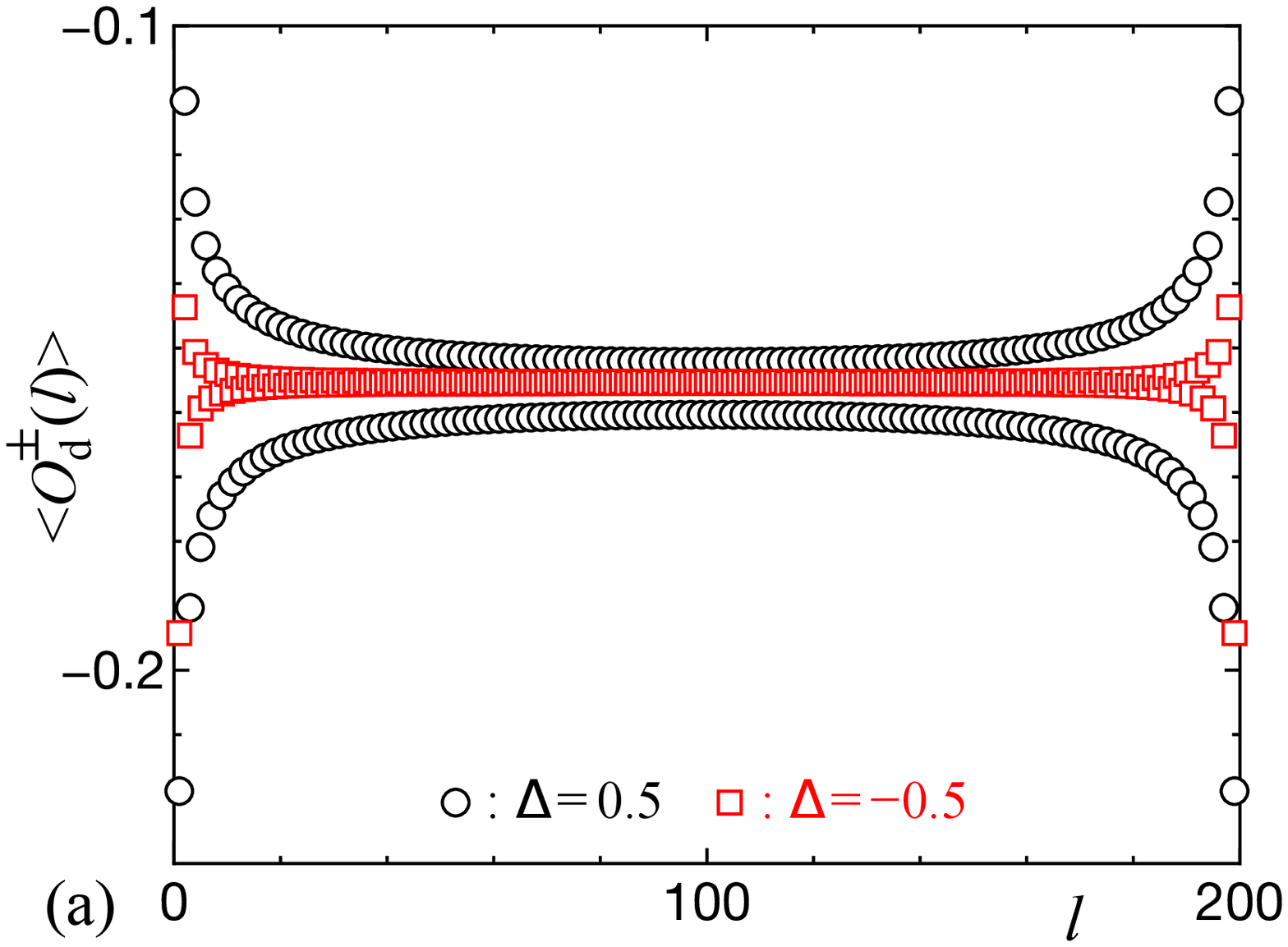}
\includegraphics[width=70mm]{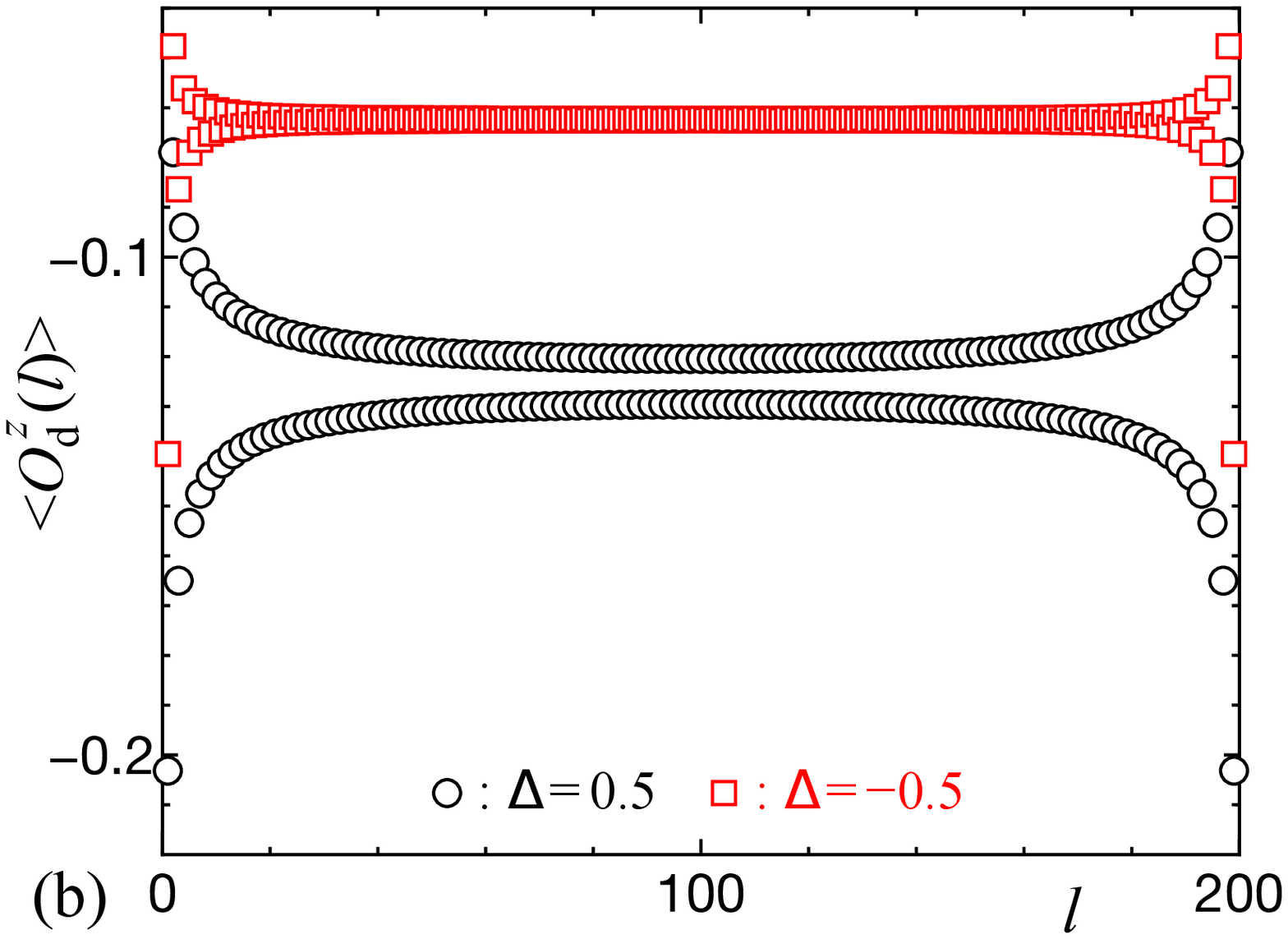}
\caption{
Expectation values of the dimer operators
(a) $\langle \mathcal{O}_{\rm d}^\pm (l) \rangle$ and
(b) $\langle \mathcal{O}_{\rm d}^z (l) \rangle$ in the ground state
of the XXZ chain (\ref{eq:HamXXZ}) for $\Delta=0.5, -0.5$,
zero magnetic field $h=0$, and $L=200$.
}
\label{fig:Od_zerofield}
\end{center}
\end{figure}

\begin{figure}
\begin{center}
\includegraphics[width=70mm]{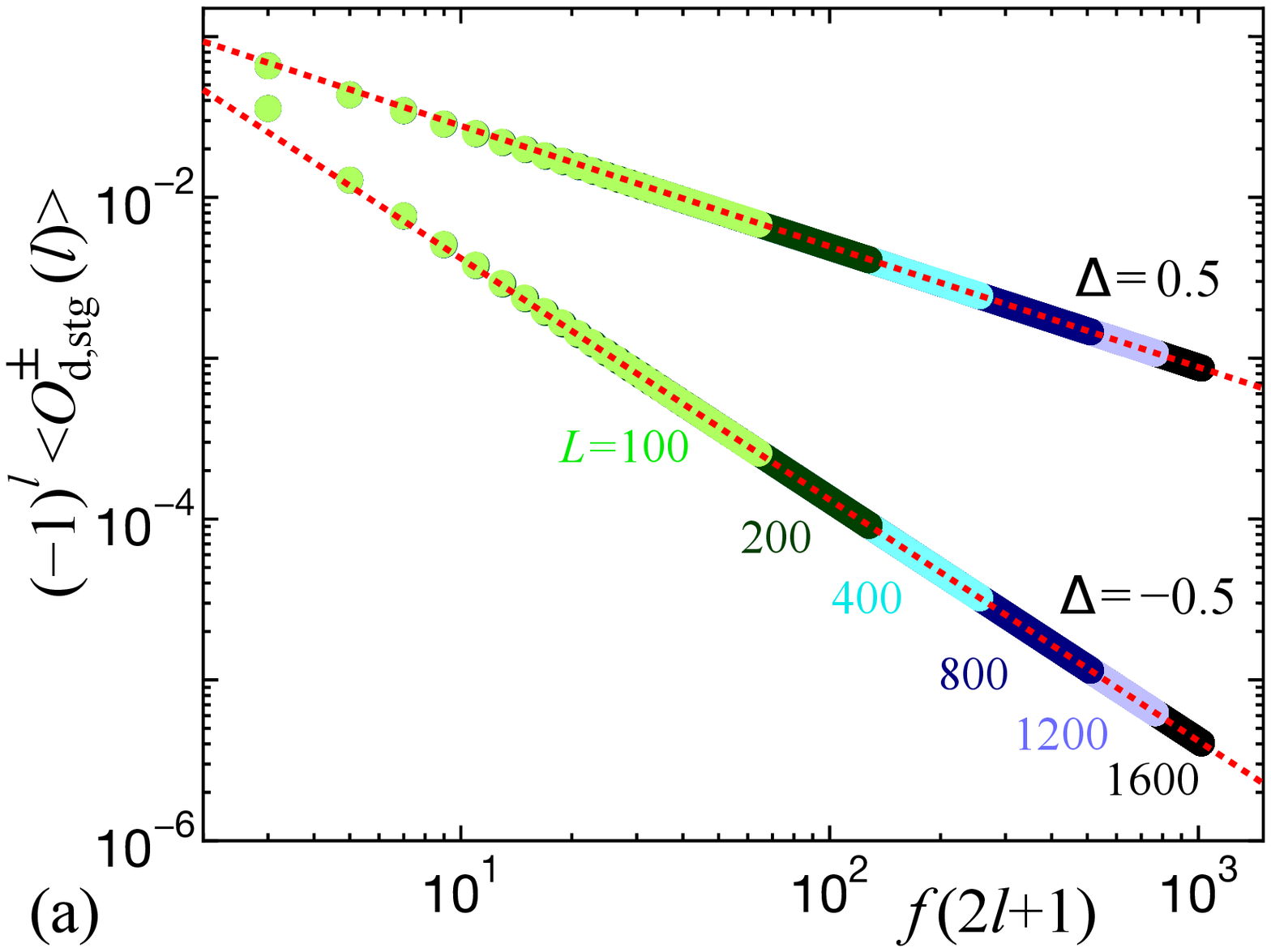}
\includegraphics[width=70mm]{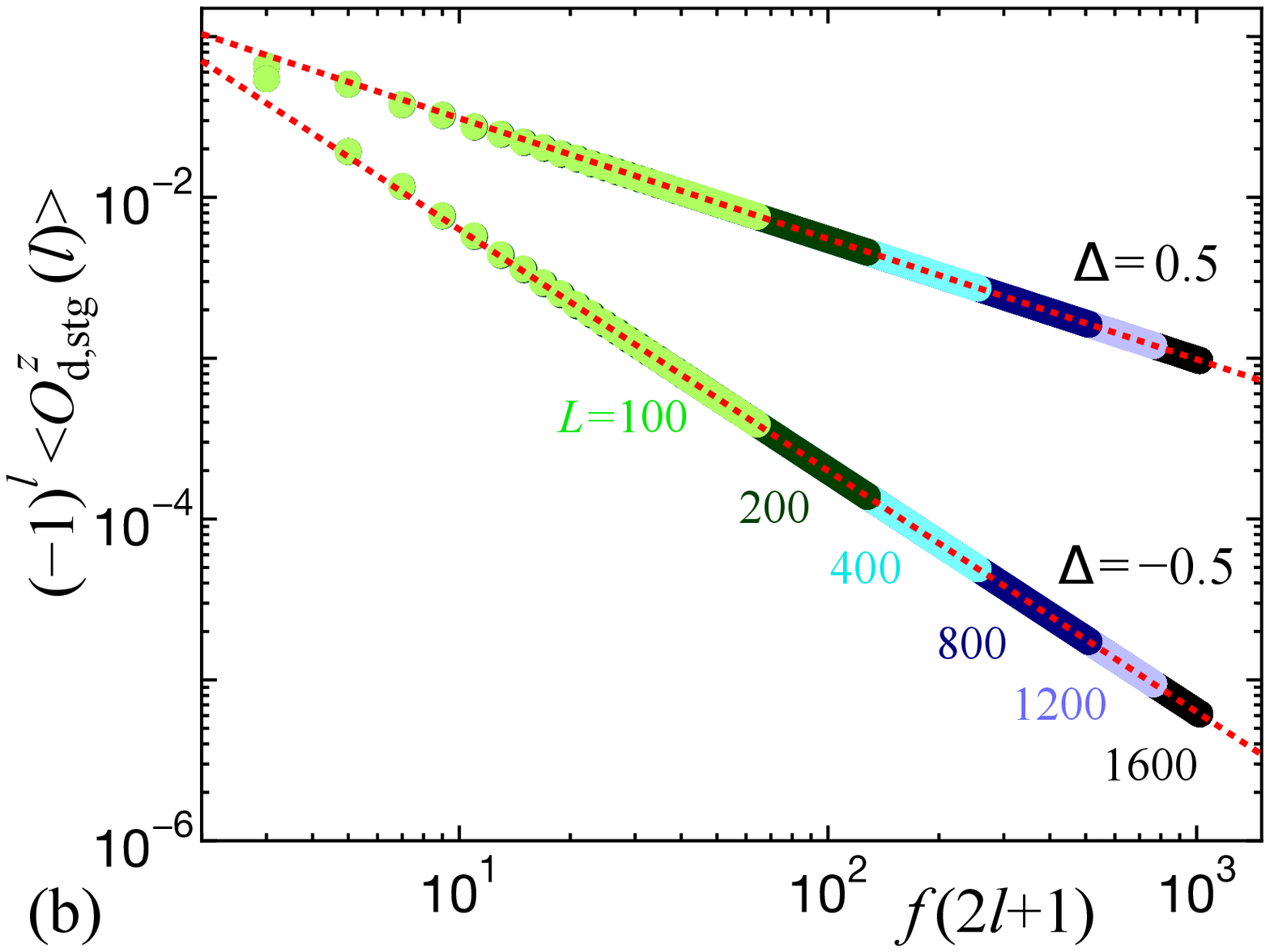}
\caption{
Staggered part of the expectation value of the dimer operators
(a) $(-1)^l \langle \mathcal{O}_{\rm d,stg}^\pm (l) \rangle$ and
(b) $(-1)^l \langle \mathcal{O}_{\rm d,stg}^z (l) \rangle$
in the ground state of the XXZ chain (\ref{eq:HamXXZ})
for $\Delta=0.5, -0.5$ and zero magnetic field $h=0$.
The data for $L=100, 200, 400, 800, 1200$, and $1600$ are plotted.
The dashed lines represent the expected behavior, $c_1^a/[f(2l+1)]^{1/2\eta}$,
with $c_1^a$ obtained by the procedure explained
in the text and $\eta$ given by Eq.\ (\ref{eq:eta_zeroM}).
}
\label{fig:Odstg_zerofield}
\end{center}
\end{figure}

In order to estimate the coefficients $c_1^a$ ($a=\pm,z$)
for the XXZ chain at zero field,
we computed the ground-state expectation values of the dimer operators,
$\langle \mathcal{O}_{\rm d}^a (l) \rangle$,
for the systems up to $L=1600$ spins.
Figure\ \ref{fig:Od_zerofield} shows the numerical results for $\Delta=0.5$
and $\Delta=-0.5$.
(Here we plot the results obtained for a rather small system size
$L=200$ for clarity.)
The ground-state expectation values of the dimer operators exhibit
sizable Friedel oscillations near open boundaries.
The staggered part of the expectation values of the dimer operators,
$\langle \mathcal{O}_{\rm d, stg}^a (l) \rangle$,
can be obtained from $\langle \mathcal{O}_{\rm d}^a (l) \rangle$
by subtracting the non-oscillating contributions,
\begin{align}
\langle \mathcal{O}_{\rm d, stg}^a (l) \rangle = 
\langle \mathcal{O}_{\rm d}^a (l) \rangle
- 
c_0^a 
+ \frac{\pi^2 c_2^a}{12(L+1)^2}
+ \frac{\bar{c}_2^a}{[f(2l+1)]^2},
\label{eq:Odstg_num}
\end{align}
where the exact values given in Eqs.\ (\ref{eq:c0_zero_field}) and
(\ref{eq:c2_and_barc2_zero_field}) are substituted for the coefficients
$c_0^a$, $c_2^a$, and $\bar{c}_2^a$.
The staggered part $\langle \mathcal{O}_{\rm d, stg}^a (l) \rangle$
obtained in this way is shown in Fig.\ \ref{fig:Odstg_zerofield}.
We see that data points of
$(-1)^l\langle \mathcal{O}_{\rm d, stg}^a (l) \rangle$ computed for different
system sizes collapse onto a single line in the log-log plot,
which corresponds to the power-law behavior
$(-1)^l\langle \mathcal{O}_{\rm d, stg}^a (l) \rangle
=c_1^a/[f(2l+1)]^{1/2\eta}$.
This demonstrates the validity of Eq.\ (\ref{eq:Odim-finite})
and indicates that the higher-order terms neglected there are indeed very small.

\begin{figure}
\begin{center}
\includegraphics[width=70mm]{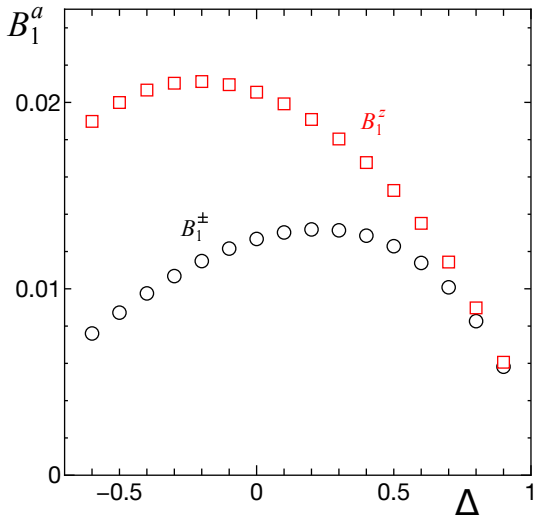}
\caption{
Amplitudes $B_1^a=(c_1^a)^2/2$ of the leading staggered term
in the dimer correlation functions, Eq.\ (\ref{eq:Odim-cor}),
in the XXZ chain (\ref{eq:HamXXZ}) at zero magnetic field $h=0$.
}
\label{fig:B1a_zerofield}
\end{center}
\end{figure}

The coefficients $c_1^a$ are obtained from
$\langle \mathcal{O}_{\rm d, stg}^a (l) \rangle$
as follows.
For an open spin chain of $L$ sites,
we calculate
\begin{eqnarray}
c_1^a(l,L)=
(-1)^l \langle\mathcal{O}_{\rm d, stg}^a (l)\rangle[f(2l+1)]^{1/2\eta}
\label{eq:c1_from_Odstg}
\end{eqnarray}
for each $l$ in the central region ($L/2-10 \le l \le L/2+10$)
and the spatial average of $c_1^a(l,L)$ over the central region
is denoted by $c_1^a(L)$.
We calculate $c_1^a(L)$ for several values of $L$
and obtain a set of data
$\mathcal{C}_1^a=\{c_1^a(L)|L=100, 200,\ldots,1600\}$.
For three different subsets of $\mathcal{C}_1^a$
we fit $c_1^a(L)$ to the polynomial
$c_1^a(L)=c_1^a(\infty) + \beta_1^a/L + \beta_2^a/L^2$;
this defines the extrapolated value $c_1^a(\infty)$
for each subset of $\mathcal{C}_1^a$.
We take the average of these $c_1^a(\infty)$
as the final estimate of $c_1^a$.
The error is determined from the largest of the differences
of the final estimate $c_1^a$ from
the extrapolated values $c_1^a(\infty)$ for the subsets of $\mathcal{C}_1^a$
and from the estimates $c_1^a(l,L)$ for the central region of the largest system $L=1600$.
In this way we have determined the coefficients $c_1^a$ for $\Delta \ge -0.6$,
but we could not obtain accurate results for $\Delta \le -0.7$, 
where the Friedel oscillations in $\langle \mathcal{O}_{\rm d}^a (l) \rangle$ 
decay so rapidly that the amplitude of oscillations
away from the boundaries becomes almost comparable to the numerical accuracy
of our DMRG data.
The results for the amplitudes $B_1^a=(c_1^a)^2/2$
of the leading staggered term of the dimer correlation functions
are presented
in Table\ \ref{tab:B1a_zerofield} and Fig.~\ref{fig:B1a_zerofield}.

\begin{table}
\caption{
Amplitudes $B_1^a=(c_1^a)^2/2$ of the leading staggered term
in the dimer correlation functions (\ref{eq:Odim-cor})
in the XXZ chain (\ref{eq:HamXXZ}) for $h=0$
as functions of the anisotropy parameter $\Delta$.
The number in the parentheses for the value of $B_1^a$
denotes the error in the last digit.
The error was estimated as described below
Eq.\ (\ref{eq:c1_from_Odstg}).\cite{systematic_error}
}
\label{tab:B1a_zerofield}
\begin{center}
\begin{tabular}{rll}
\hline
\hline
 $\Delta$ & $B_1^\pm$ & $B_1^z$ \\
\hline
  0.9 &   0.00582(6) &   0.00606(7) \\
  0.8 &   0.00826    &   0.00898    \\
  0.7 &   0.01008(2) &   0.01144(2) \\
  0.6 &   0.01139(2) &   0.01351(3) \\
  0.5 &   0.01229(2) &   0.01528(3) \\
  0.4 &   0.01285(1) &   0.01677(2) \\
  0.3 &   0.01314(1) &   0.01804(1) \\
  0.2 &   0.01319(1) &   0.01909(1) \\
  0.1 &   0.01302(1) &   0.01992(2) \\
  0.0 &   0.01267(1) &   0.02055(2) \\
 $-0.1$ &   0.01216(1) &   0.02095(2) \\
 $-0.2$ &   0.01149(1) &   0.02112(2) \\
 $-0.3$ &   0.01068(1) &   0.02103(2) \\
 $-0.4$ &   0.00975(1) &   0.02066(2) \\
 $-0.5$ &   0.00872(1) &   0.01999(3) \\
 $-0.6$ &   0.00760(7) &   0.0190(3) \\
\hline
\hline
\end{tabular}
\end{center}
\end{table}

\subsection{Application}\label{subsec:appl_zeroM}

The high-precision data of the coefficients $c_1^a$ can be used
for quantitative analysis of physical quantities related to
the dimer operators,
including spin-Peierls instability, dynamical structure factors of dimer
correlations, and interchain dimer-dimer couplings in quasi-1D systems.
As an example of such applications, we discuss the excitation gap
in the XXZ chain with bond alternation in this section.

Let us consider the bond-alternating spin-1/2 XXZ chain,
whose Hamiltonian is
\begin{equation}
\mathcal{H}_{\rm ba} =
J \sum_{l=1}^{L-1} \left[ 1-(-1)^l \delta \right] \!
( S^x_l S^x_{l+1} + S^y_l S^y_{l+1} + \Delta S^z_l S^z_{l+1} ),
\label{eq:HamXXZ_ba}
\end{equation}
where $\delta$ is a positive parameter controlling the magnitude of
the bond alternation.
We assume the easy-plane anisotropy, $|\Delta| < 1$.
From Eq.\ (\ref{eq:O_boson_formula}), it is found that the low-energy
effective Hamiltonian for Eq.\ (\ref{eq:HamXXZ_ba}) is given by
\begin{equation}
\widetilde{\mathcal{H}}_{\rm ba} = \widetilde{\mathcal{H}}_0
- J \delta \!\left( 2 c_1^\pm + \Delta c_1^z \right) \!
  \int \! dx \cos[\sqrt{2\pi}\phi(x)] + \cdots,
\label{eq:tildeHamXXZ_ba}
\end{equation}
where $\widetilde{\mathcal{H}}_0$ is the Gaussian model in
Eq.\ (\ref{eq:Gaussian}).
Since the nonlinear term $\cos[\sqrt{2\pi}\phi(x)]$ has a scaling dimension
$1/(2\eta)$ at the Gaussian fixed point, it is a relevant perturbation and
opens an excitation gap if $\eta > 1/4$
(i.e., $\Delta > -1/\sqrt{2}$).
In this case the excitation gap $E_{\rm g}(\delta)$
for small bond alternation $\delta \ll 1$ is given by\cite{Zamolodchikov1995}
\begin{equation}
\frac{E_{\rm g}(\delta)}{J}
= A(\Delta) \! \left| \delta(2c_1^\pm+\Delta c_1^z) \right|^{2\eta/(4\eta-1)}
\label{eq:dimer_gap}
\end{equation}
with
\begin{eqnarray}
A(\Delta) = \frac{2 v}{\sqrt{\pi} J}
\frac{\Gamma\bigl(\frac{1}{8\eta-2}\bigr)}
     {\Gamma\bigl(\frac{2\eta}{4\eta-1}\bigr)}
\!\left[
\frac{\pi J}{2v}
\frac{\Gamma\bigl(1-\frac{1}{4\eta}\bigr)}
     {\Gamma\bigl(\frac{1}{4\eta}\bigr)}
\right]^{2\eta/(4\eta-1)}.
\label{eq:dimer_gap_coeff}
\end{eqnarray}
Note that the parameter $\eta$ and the spin velocity $v$ are
functions of $\Delta$
[see Eqs.\ (\ref{eq:eta_zeroM}) and (\ref{eq:v_zero_field})].
Thus, with the estimates of $c_1^a$ obtained
in Sec.~\ref{subsec:numerics_zeroM},
we can determine the excitation gap from Eqs.\ (\ref{eq:dimer_gap})
and (\ref{eq:dimer_gap_coeff})
without any free parameter.

To confirm this theory, we numerically calculated the excitation gap
$E_{\rm g}(\delta)$ for $\Delta=0.5$ and $\delta = 2^{-3}, ..., 2^{-10}$
using the DMRG method.
The gap $E_{\rm g}(\delta)$ was obtained as follows.
We first calculated the excitation gap for finite open spin chains
of various lengths up to $L=3200$,
using the relation
\begin{eqnarray}
E_{\rm g}(\delta, L)=E_0(\delta; L,1) - E_0(\delta; L,0),
\label{eq:gap_num}
\end{eqnarray}
where $E_0(\delta; L, S^z_{\rm tot})$ is the lowest energy in the subspace
in which the total magnetization $\sum_l S^z_l = S^z_{\rm tot}$.
We thus obtained a set of data
$\mathcal{E}=\{E_{\rm g}(\delta,L)|L=100, 200, \ldots, 3200\}$.
For three different subsets of $\mathcal{E}$
we fit $E_{\rm g}(\delta, L)$
to a second-order polynomial,
$E_{\rm g}(\delta, L)=E_{\rm g}(\delta, \infty)+\beta_1/L+\beta_2/L^2$,
to obtain the extrapolated value $E_{\rm g}(\delta, \infty)$ 
for each subset of $\mathcal{E}$.
We took the average of $E_{\rm g}(\delta, \infty)$ for the subsets
as the final estimate of $E_{\rm g}(\delta)$.
The error in $E_{\rm g}(\delta)$, which is estimated from the 
difference between the final estimate $E_{\rm g}(\delta)$
and the extrapolation $E_{\rm g}(\delta, \infty)$
for the subsets of $\mathcal{E}$, 
is less than $3.9 \times 10^{-5}J$.

\begin{figure}
\begin{center}
\includegraphics[width=70mm]{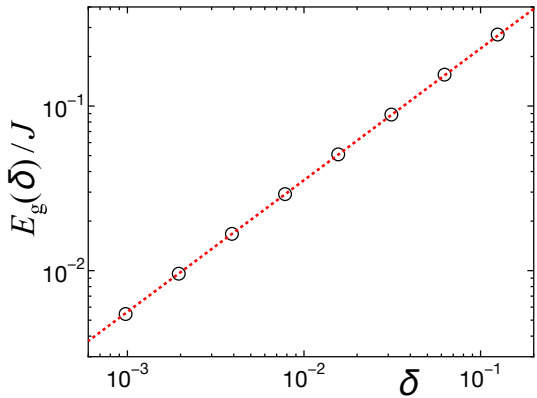}
\caption{
Excitation gap $E_{\rm g}(\delta)$ in the bond-alternating XXZ chain
(\ref{eq:HamXXZ_ba}) at $\Delta=0.5$.
The circles represent the gap $E_{\rm g}(\delta)$ obtained using
the DMRG method and the extrapolation
as described below Eq.\ (\ref{eq:gap_num}).
The dotted line shows the theoretical curve from Eqs.\ (\ref{eq:dimer_gap})
and (\ref{eq:dimer_gap_coeff}), in which the exact values of $\eta$
and $v$ [Eqs.\ (\ref{eq:eta_zeroM}) and (\ref{eq:v_zero_field})] and
the coefficients $c_1^\pm$ and $c_1^z$ obtained
in Sec.\ \ref{subsec:numerics_zeroM} are substituted.
}
\label{fig:gap_d05}
\end{center}
\end{figure}

In Fig.\ \ref{fig:gap_d05}, we show $E_\mathrm{g}(\delta)$, together with
a plot of Eq.\ (\ref{eq:dimer_gap}) calculated with $c_1^a$ obtained
in the previous section.
Clearly, the numerical and analytic results are in excellent
agreement,\cite{diff_gap_d05}
demonstrating the accuracy of the estimates of $c_1^a$ and
the validity of the theory.

\section{SU(2) symmetric case}\label{sec:SU2}

In this section we discuss the SU(2) symmetric case where
$\Delta = 1$ and $h=0$ in Eq.\ (\ref{eq:HamXXZ}).
In this case the marginally irrelevant operator
in the low-energy effective theory brings about logarithmic corrections
in various physical quantities.\cite{Giamarchi-text,GogolinNT-text,Affleck-text,Lukyanov1998,VekuaS2016,Affleck1998}
For example, the leading behavior of the dimer correlation function is
\begin{eqnarray}
&&\langle\mathcal{O}_\mathrm{d}^a(l)\mathcal{O}_\mathrm{d}^a(l+r)\rangle
-\langle\mathcal{O}_\mathrm{d}^a(l)\rangle
 \langle\mathcal{O}_\mathrm{d}^a(l+r)\rangle
\nonumber \\
&&~~~~~~~~
=\widetilde{B}_1\frac{(-1)^r}{r\,(\ln r)^{3/2}}+\cdots,
\end{eqnarray}
for $r\gg1$, where $\widetilde{B}_1$ is a constant common to $a = \pm, z$.
This behavior can be understood within the scheme of the previous section
as follows;
in the SU(2) symmetric limit,
the correlation amplitude $B_1=(c_1)^2/2$ is renormalized and
acquires logarithmic dependence on the length or energy scale of interest.
Namely, $B_1\propto(\ln r)^{-3/2}$ and
$c_1\propto(\ln r)^{-3/4}$.
In the following, we reversely employ the analysis of 
Sec.\ \ref{subsec:appl_zeroM}; that is, we deduce the amplitude $\widetilde{B}_1$
from the dependence of the excitation gap $E_\mathrm{g}$
on the bond alternation $\delta$.

Let us consider the Heisenberg spin chain with the bond alternation
[Eq.\ (\ref{eq:HamXXZ_ba}) with $\Delta=1$].
The low-energy effective Hamiltonian is written in terms of the bosonic fields
as
\begin{align}
\widetilde{\mathcal{H}}_{\rm ba,SU(2)} =& \,
\widetilde{\mathcal{H}}_0
- 3 c_1 \delta J \int dx \cos[\sqrt{2\pi}\phi(x)]
\nonumber \\
&+ g \int dx \cos[\sqrt{8\pi}\phi(x)] + \cdots,
\label{eq:tildeHamXXZ_ba_SU2}
\end{align}
where $\widetilde{\mathcal{H}}_0$ is the Gaussian model
in Eq.\ (\ref{eq:Gaussian}) and $c_1=c_1^\pm=c_1^z$.
It is important to note that we have included the marginally irrelevant term,
$g \int dx \cos[\sqrt{8\pi}\phi(x)]$, in the effective Hamiltonian.
In the absence of the bond alternation ($\delta=0$),
the coupling constant $g$ is renormalized to zero
as $g\sim [\ln(J/E)]^{-1}$ with decreasing energy scale $E$.
When the bond alternation is present, $\delta \ne 0$,
the renormalization of the coupling constant $g$ is stopped
at the energy scale of the excitation gap $E_{\rm g}$,
where $g$ takes a finite value.
Using the renormalization-group scheme from Ref.\ \onlinecite{Lukyanov1998},
the relation between the gap $E_\mathrm{g}$ and
the running coupling constant $g$ can be chosen as
\begin{eqnarray}
\frac{E_{\rm g}}{J} =
\sqrt{2\pi^3}\, e^{\gamma_{\rm E}} g^{-1/2} e^{-1/g},
\label{eq:Eg-g}
\end{eqnarray}
where $\gamma_{\rm E} \simeq 0.5772...$ is the Euler constant.

We suppose that the gap formula of Eqs.\ (\ref{eq:dimer_gap}) and
(\ref{eq:dimer_gap_coeff}) holds also in the SU(2) symmetric case
and that logarithmic corrections manifest themselves
through the renormalized coefficient $c_1$.
Thus, we substitute $\eta=1$ and $v=\pi J/2$, which are the fixed-point values
in the SU(2) case in the absence of the bond alternation, into
Eqs.\ (\ref{eq:dimer_gap}) and (\ref{eq:dimer_gap_coeff}).
Then we write
\begin{equation}
c_1 = \frac{1}{(2\pi^3)^{1/4}} \frac{g^{3/4}}{C(g)},
\label{eq:c1-g-Cg}
\end{equation}
where
\begin{equation}
C(g) =
(2\pi^3)^{-1/4} g^{3/4}
\frac{3\delta\Gamma\left(\frac{3}{4}\right)}{\Gamma\left(\frac{1}{4}\right)}
\!
\left[
\frac{\Gamma\left(\frac{2}{3}\right)}
     {\sqrt{\pi}\,\Gamma\left(\frac{1}{6}\right)}
\frac{E_{\rm g}}{J}
\right]^{-3/2}.
\label{eq:Cg}
\end{equation}
We have defined $C(g)$ in such a way that the prefactor
$g^{3/4}$ in Eq.\ (\ref{eq:c1-g-Cg}) incorporates
the scaling $c_1\propto g^{3/4}$ at $g \ll 1$.
It is then natural to expect that $C(g)$ should be expanded
in powers of $g$, 
\begin{eqnarray}
C(g) = C_0 + C_1 g + C_2 g^2 + \cdots
\label{eq:Cg_poly}
\end{eqnarray}
for $g\ll1$.

\begin{figure}
\begin{center}
\includegraphics[width=70mm]{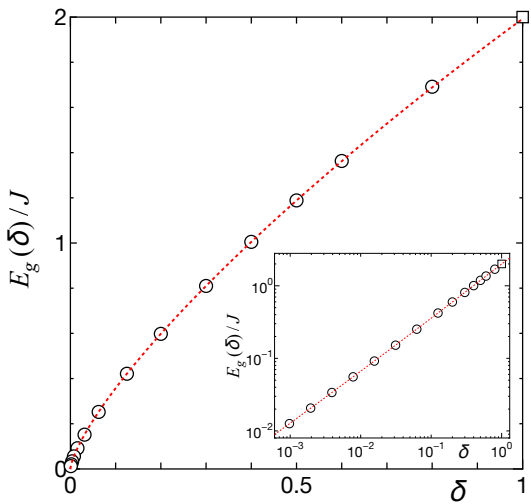}
\caption{
Excitation gap $E_{\rm g}(\delta)$ in the bond-alternating Heisenberg chain,
Eq.\ (\ref{eq:HamXXZ_ba}) with $\Delta=1$.
The circles represent the numerical data extrapolated to the thermodynamic
limit $L \to \infty$, and the square is the exact value $E_{\rm g}(\delta=1)=2$.
The red dotted line is the theoretical curve,
Eqs.\ (\ref{eq:Eg-g}) and (\ref{eq:delta-Eg-g}), with $C_2 = 1.80$.
The inset shows the same figure in a log-log scale.
}
\label{fig:gap_d10}
\end{center}
\end{figure}

In order to estimate the constants $C_0$, $C_1$, and $C_2$
in Eq.\ (\ref{eq:Cg_poly}), we calculated numerically the excitation gap
$E_{\rm g}(\delta)$ in the bond-alternating chain (\ref{eq:HamXXZ_ba})
with $\Delta=1$ and $\delta = 2^{-10}, ..., 2^{-3}, 0.2, ..., 0.8$
using the DMRG method.
Previous works have obtained the excitation gap $E_{\rm g}(\delta,L)$
for $L \lesssim 200$ spins.\cite{PapenbrockBDSS2003,KumarRSS2007}
Here, we computed $E_{\rm g}(\delta,L)$ for the finite open chains 
up to $L \le 3200$ ($L \le 800$) spins 
with $2^{-10} \le \delta \le 2^{-3}$ ($0.2 \le \delta \le 0.8$).
We then extrapolated the data to $L \to \infty$ in the same manner 
as in Sec.\ \ref{subsec:appl_zeroM} and obtained the estimate of 
the gap $E_{\rm g}(\delta)$ in the thermodynamic limit.
The error in $E_{\rm g}(\delta)$ is estimated to be 
less than $1.5 \times 10^{-5}J$.
The numerical results for $E_{\rm g}(\delta)$
are shown by open circles in Fig.~\ref{fig:gap_d10}.

\begin{figure}
\begin{center}
\includegraphics[width=70mm]{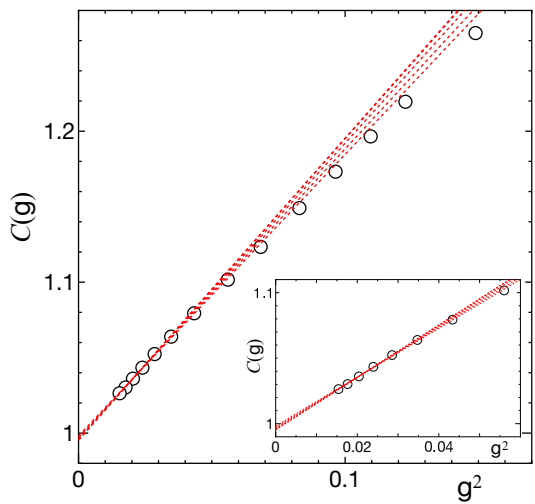}
\caption{
$C(g)$ as a function of $g^2$.
The circles represent the numerical data.
The red dotted lines show the fitting to
$C(g) = C_0 + C_2 g^2$
of the data points at the $n$ smallest $g$ ($n=4,5,...,8$).
The inset shows the same figure on an enlarged scale.
}
\label{fig:Cg-g}
\end{center}
\end{figure}

Having determined $E_\mathrm{g}(\delta)$ numerically,
we use Eq.\ (\ref{eq:Eg-g}) to obtain the renormalized coupling constant
$g$ as a function of $\delta$.
Then we substitute $E_\mathrm{g}(\delta)$ and $g(\delta)$
into the right-hand side of Eq.\ (\ref{eq:Cg}) to obtain $C(g)$
for each $\delta$ calculated.
In Fig.\ \ref{fig:Cg-g}, we plot the so-obtained $C(g)$
(open circles).
As clearly shown in Fig.\ \ref{fig:Cg-g},
when plotted as a function of $g^2$,
$C(g)$ exhibits a linear behavior and approaches unity as $g^2 \to 0$.
Fitting $C(g)$ of the $n$-smallest $g$ ($n=4 - 8$)
to Eq.\ (\ref{eq:Cg_poly}) while assuming $C_1=0$ and neglecting
the higher-order terms $\mathcal{O}(g^3)$, we obtain $0.995 \le C_0 \le 0.998$.
These results indicate that $C_0=1$ and $C_1=0$.
Then fitting $C(g)$ while assuming $C_0=1$ and $C_1=0$ yields
$C_2 \simeq 1.80(3)$.

The results obtained above lead to the following expression
for the excitation gap.
From Eqs.\ (\ref{eq:Cg}) and (\ref{eq:Cg_poly}), we can write
the bond alternation $\delta$ in terms of $E_{\rm g}$ and $g$ as
\begin{equation}
\delta =
\frac{2 \Gamma\left(\frac{1}{4}\right)}
     {3\Gamma\left(\frac{3}{4}\right)} \!
\left[\frac{\Gamma\left(\frac{2}{3}\right)}
           {\sqrt{2}\Gamma\left(\frac{1}{6}\right)}\frac{E_{\rm g}}{J}
\right]^{3/2}
g^{-3/4} \bigl( 1 + C_2 g^2 \bigr),
\label{eq:delta-Eg-g}
\end{equation}
where we have substituted $C_0=1$ and $C_1=0$ in Eq.\ (\ref{eq:Cg_poly})
and omitted the higher-order terms $\mathcal{O}(g^3)$ in $C(g)$.
Equations (\ref{eq:Eg-g}) and (\ref{eq:delta-Eg-g}) give a parametric
representation of $E_{\rm g}(\delta)$
in terms of $g$.
In Fig.\ \ref{fig:gap_d10}, we plot the gap $E_{\rm g}(\delta)$ calculated
from Eqs.\ (\ref{eq:Eg-g}) and (\ref{eq:delta-Eg-g}).
Clearly, the theoretical curve reproduces the numerical data.
We emphasize that the agreement between the theory and numerical data is
excellent even at the large bond alternation, $\delta \to 1$, suggesting 
that the effect of the higher-order terms $\mathcal{O}(g^3)$ in $C(g)$ 
on the excitation gap $E_{\rm g}(\delta)$ is negligible.
Our theory with Eqs.\ (\ref{eq:Eg-g}) and (\ref{eq:delta-Eg-g}) thereby 
provides accurate values of $E_{\rm g}(\delta)$
for the whole range of the bond alternation $0 < \delta \le 1$.

In addition, the above theory allows us to derive the long-distance
behavior of the dimer correlation function in the uniform Heisenberg chain
[Eq.\ (\ref{eq:HamXXZ}) with $\Delta = 1$] in zero field $h=0$.
Substituting Eq.\ (\ref{eq:c1-g-Cg}) with $C(g)=1$ into
Eq.\ (\ref{eq:Odim-cor}) with $B_1^a = (c_1^a)^2/2$ and
replacing $g$ by $(\ln r)^{-1}$, we obtain
\begin{eqnarray}
&&\langle\mathcal{O}_{\rm d}^a(l)\mathcal{O}_{\rm d}^a(l+r)\rangle
-\langle\mathcal{O}_{\rm d}^a(l)\rangle
 \langle\mathcal{O}_{\rm d}^a(l+r)\rangle
\nonumber \\
&&~~~~~~~~
= \frac{1}{(2\pi)^{3/2}} \frac{(-1)^r}{r\,(\ln r)^{3/2}} + \cdots,
\label{eq:Odim-cor_SU2}
\end{eqnarray}
where $a=\pm, z$ (no summation is taken for the repeated index $a$).
Note that the correlation functions of $\mathcal{O}_{\rm d}^\pm$ and
$\mathcal{O}_{\rm d}^z$ are identical due to the SU(2) symmetry.
We note that the amplitude $\widetilde{B}_1=(2\pi)^{-3/2} = 0.0635...$ is
in good agreement with the recent numerical estimate $0.067$ reported in
Ref.\ \onlinecite{VekuaS2016}.

\section{XXZ chain with nonzero magnetization}\label{sec:finiteM}

\subsection{Theory}\label{subsec:XXZ_theory_finiteM}

In this section, we study the XXZ chain (\ref{eq:HamXXZ}) in the TLL phase
with a partial spin polarization
under finite external field $h_{\rm c} < h < h_{\rm s}$.
Here, $h_{\rm c}$ is the lower critical field
($h_{\rm c}=0$ for $-1 < \Delta \le 1$ and $h_{\rm c} > 0$ for $\Delta > 1$),
while $h_{\rm s}=J(1+\Delta)$ is the saturation field.
The low-energy effective theory in this case is
the Gaussian model (\ref{eq:Gaussian}) again.
In the partially polarized state with $0 < M < 1/2$,
the Fermi momentum $k_{\rm F}$ of the Jordan-Wigner fermions is shifted
from the commensurate value $k_{\rm F}=\pi/2$ at $M=0$
to the incommensurate one $k_{\rm F} = \pi(\frac{1}{2}+M)$.
The boson-field expression of the dimer operator (\ref{eq:Odim})
is then modified from Eq.\ (\ref{eq:O_boson_formula}) into
\begin{align}
\mathcal{O}_{\rm d}^a(l)=&\,
c_0^a + c_1^a(-1)^l\cos[Q x_l + \sqrt{2\pi}\phi(x_l)]
\nonumber\\
& + c_\phi^a \left(\frac{d\phi(x_l)}{dx}\right)^2
+ c_\theta^a \left(\frac{d\theta(x_l)}{dx}\right)^2
\nonumber\\
& + c_g^a \cos[2Qx_l + \sqrt{8\pi}\phi(x_l)]
+\cdots
\label{eq:O_boson_formula_field}
\end{align}
for $a=\pm,z$.
The wave number $Q$ of the leading oscillating term is 
$Q=2\pi M$ in the limit $L\to\infty$.

In the same manner as in Sec.\ \ref{subsec:XXZ_theory},
we can calculate the ground-state expectation values of the dimer operators
in Eq.\ (\ref{eq:Odim}) in finite chains with open boundaries.
For the partially polarized state, we find it necessary
to optimize the positions at which the Dirichlet boundary condition
is imposed, in order to achieve a better fitting of the numerical
data.\cite{Fath2003,HikiharaMFK2010} 
We thus employ
the Dirichlet boundary conditions $\phi(x_0) = \phi(L+1-x_0) = 0$, 
instead of Eq.\ (\ref{eq:Dirichlet}).
Accordingly, the one-point functions of the dimer operators become
\begin{align}
\langle \mathcal{O}_{\rm d}^a (l) \rangle =&\,
c_0^a
+ \frac{c_1^a (-1)^l \cos[\tilde{Q}(l+1/2-x_0)]}
       {[\tilde{f}(2l+1-2x_0)]^{1/2\eta}} 
\nonumber \\
& - \frac{\pi^2 c_2^a}{12(L+1-2x_0)^2}
 - \frac{\bar{c}_2^a}{[\tilde{f}(2l+1-2x_0)]^2}
\nonumber \\
& + \frac{c_g^a\cos[2\tilde{Q}(l+1/2-x_0)]}{[\tilde{f}(2l+1-2x_0)]^{2/\eta}}
 + \cdots,
\label{eq:Odim-finite_finiteM}
\end{align}
where $\tilde{Q} = 2\pi ML/(L+1-2x_0)$ and 
\begin{equation}
\tilde{f}(x) =
\frac{2(L+1-2x_0)}{\pi} \sin\!\left(\frac{\pi |x|}{2(L+1-2x_0)}\right).
\label{eq:tildef}
\end{equation}
The parameter $\eta$ can be determined exactly by solving
the integral equations obtained from the Bethe ansatz.\cite{BogoliubovIK1986,QinFYOA1997,CabraHP1998}
We have kept the last term ($\propto c_g^a$) in
Eq.\ (\ref{eq:Odim-finite_finiteM}) 
since it becomes larger than the third and fourth terms
for $\eta>1$, which realizes at $\Delta>1$ and not too large $M>0$.

The coefficients of the uniform parts, $c_0^a$, $c_2^a$, $\bar{c}_2^a$,
and $c_g^a$,
are related to the ground-state energy density $e_0$, the spin velocity $v$,
the exponent $\eta$, and the coupling constant $g$ through equations similar to
Eqs.\ (\ref{eq:c0_to_e0}), (\ref{eq:cphi_and_ctheta_to_e0}),
and (\ref{eq:cg_to_e0}),
while explicit closed formulas for $e_0$, $v$, $\eta$, and $g$ are
not available for $0<M<1/2$.
On the other hand, the exact values of the coefficients $c_1^a$ of
the oscillating terms are not known
except for the free-fermion case $\Delta = 0$,
\begin{subequations}
\label{eq:c1_XY_finiteM}
\begin{align}
c_1^\pm(\Delta=0) &= \frac{1}{2\pi},
\label{eq:c1pm_XY_finiteM} \\
c_1^z(\Delta=0) &= \frac{2}{\pi^2} [\cos(\pi M) + \pi M \sin(\pi M)].
\label{eq:c1zz_XY_finiteM}
\end{align}
\end{subequations}
We will determine the coefficients $c_1^a$
in the following numerical analysis.

\subsection{Numerical results}\label{subsec:numerics_finiteM}

Using the DMRG method, we calculated the expectation values of
the dimer operators $\langle \mathcal{O}_{\rm d}^a (l) \rangle$
in the partially-polarized ground state of the XXZ chain (\ref{eq:HamXXZ})
with $L=100, 200$, and $400$ spins for fixed magnetization $M$.
We then fit the data to the analytic form (\ref{eq:Odim-finite_finiteM})
by taking $c_1^a$, $\bar{c}_2^a$, $c_g^a$, 
$c_u^a := c_0^a - \pi^2 c_2^a/[12(L+1-2x_0)^2]$, 
and $x_0$ as fitting parameters.\cite{optimized_x0}
The exponent $\eta$ was obtained from
the Bethe ansatz integral equations.

\begin{figure}
\begin{center}
\includegraphics[width=70mm]{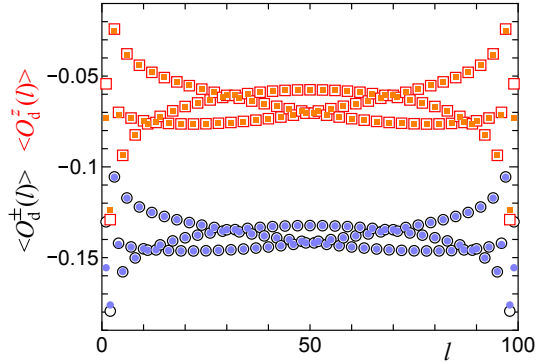}
\caption{
Expectation values of the dimer operators in the ground state of
the XXZ chain (\ref{eq:HamXXZ}) for $\Delta = 0.5$, $M=0.16$, and $L=100$.
The circles and squares correspond to
$\langle \mathcal{O}_{\rm d}^\pm (l) \rangle$
and $\langle \mathcal{O}_{\rm d}^z (l) \rangle$, respectively.
The open and solid symbols represent the DMRG data and the fitting results,
respectively.
}
\label{fig:Od_finitefield}
\end{center}
\end{figure}

We show in Fig.\ \ref{fig:Od_finitefield} the DMRG data and
the fitting results for $\Delta=0.5$ and $M=0.16$. 
(The data for the small system $L=100$ are shown for clarity.)
The agreement between the DMRG data and the fits is excellent,
which demonstrates the validity of Eq.\ (\ref{eq:Odim-finite_finiteM})
and justifies our scheme for estimating $c_1^a$.

\begin{figure}
\begin{center}
\includegraphics[width=70mm]{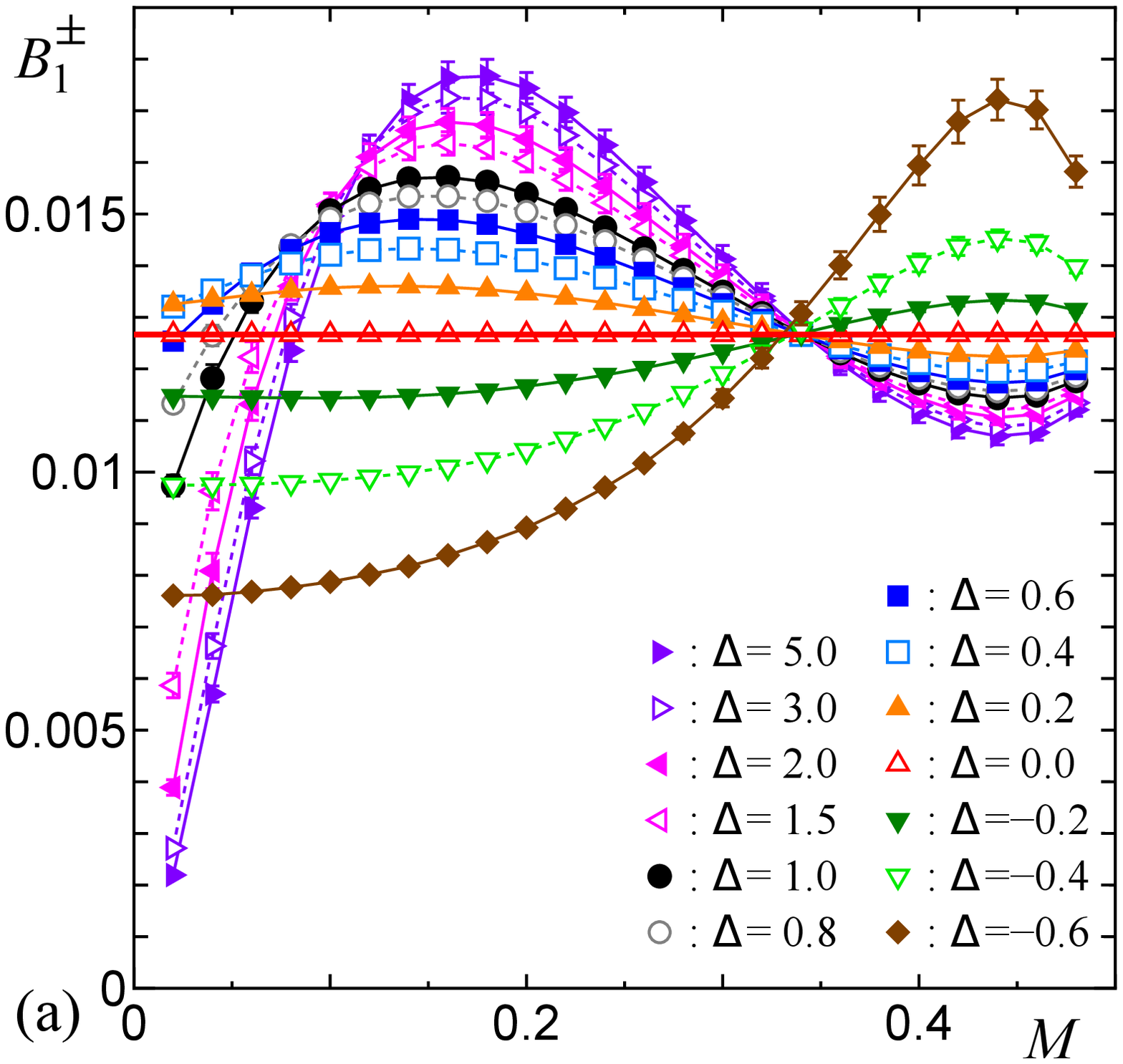}
\includegraphics[width=70mm]{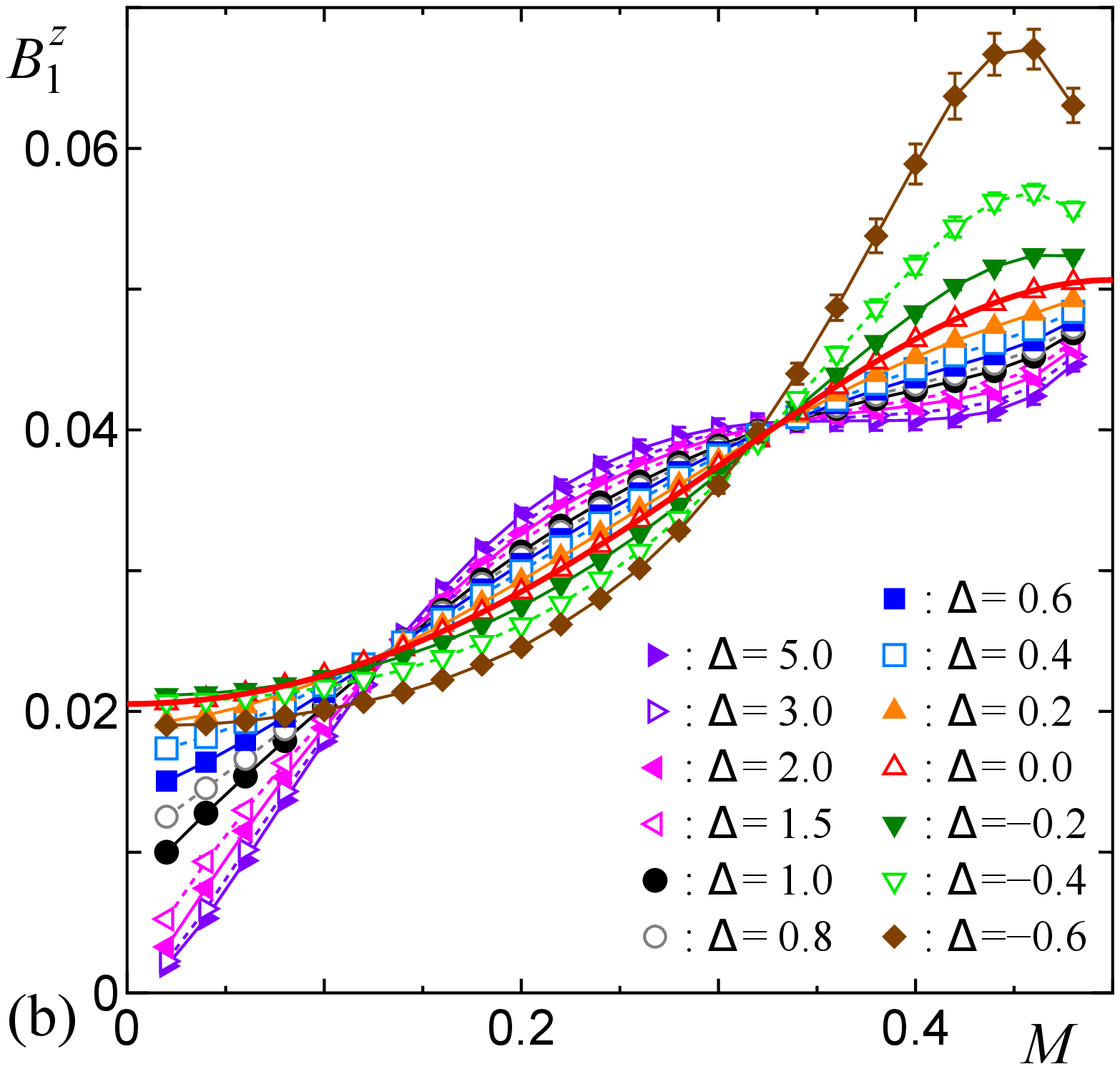}
\caption{
Amplitudes $B_1^a=(c_1^a)^2/2$ of the leading oscillating term in the dimer
correlation functions in Eq.\ (\ref{eq:Odim-cor}) as functions of $M$
for various values of the anisotropy parameter $\Delta$:
(a) $B_1^\pm$ and (b) $B_1^z$.
The thick lines represent the exact results for $\Delta = 0$,
Eq.\ (\ref{eq:c1_XY_finiteM}).
The thin lines are guides for the eye.
}
\label{fig:B1a_finitefield}
\end{center}
\end{figure}

For each system size $L$, we fit the numerical data of 
$\langle \mathcal{O}_{\rm d}^a (l) \rangle$ in three different ranges of $l$
to estimate the coefficients $c_1^a$ ($a=\pm,z$), 
which we denote $c_1^a(i,L)$ ($i=1,2,3$),
and took their averages as the estimate $c_1^a(L)$
for the system size $L$.
Then, we extrapolated the results for $L=100, 200,$ and $400$ by 
fitting them to the polynomial form
$c_1^a(L) = c_1^a(\infty) + \beta_1^a/L$
and took the extrapolated value $c_1^a(\infty)$
as the final estimate of $c_1^a$.
The error was determined from the differences between the final estimate
and the estimates $c_1^a(i,L)$ at $L=400$.
Figure\ \ref{fig:B1a_finitefield} shows the so-obtained values of
the amplitudes $B_1^a = (c_1^a)^2/2$ of the dimer correlation functions
in Eq.\ (\ref{eq:Odim-cor}).
We note that 
the numerical estimates for the free-fermion case ($\Delta=0$)
agree with the exact values in Eq.\ (\ref{eq:c1_XY_finiteM}).
Figure~\ref{fig:B1a_finitefield} also indicates that
in the saturation limit $M \to 1/2$, the amplitudes converge at
universal values,
$B_1^\pm = 1/(8\pi^2)$ and $B_1^z = 1/(2\pi^2)$.
This behavior is easily understood as the $\Delta S^z_lS^z_{l+1}$
interactions between magnons are not effective
in the limit of dilute magnon density, $M \to 1/2$.
The numerical data of the amplitudes $B_1^a$ are presented 
in the Supplemental Material.\cite{supplement}

Another interesting feature found in Fig.\ \ref{fig:B1a_finitefield} is that
the curves of $B_1^a$ for different values of $\Delta$ seem to intersect
at an intermediate value of magnetization, $M \simeq 0.33 - 0.34$.
Interestingly enough,
the amplitude $A_1^z$ of the longitudinal spin-spin
correlation function $\langle S^z_l S^z_{l+r} \rangle$
[Eq.\ (\ref{eq:SzSzcor})]
is also found\cite{HikiharaF2004} to
exhibit a similar behavior of intersection of $\Delta$-dependent curves
at $M \simeq 0.365$
[see Fig.~2(c) in Ref.\ \onlinecite{HikiharaF2004}].
At present, we do not know exactly whether and why these correlation amplitudes
really become independent of $\Delta$ at some intermediate $M$.
Furthermore, it is not clear whether or not the values of $M$
at which $B_1^a$ and $A_1^z$ become independent of $\Delta$
are the same.
These questions are open for future studies.

\section{Conclusion}\label{sec:conc}

We have studied the dimer correlation functions in the ground state of
the spin-1/2 XXZ chain in the critical Tomonaga-Luttinger-liquid regime.
We have determined with high accuracy the amplitudes of the leading
oscillating terms of the dimer correlation functions in the XXZ chain
for both zero and finite magnetic fields,
using the bosonization and DMRG methods.
We have also investigated the dimer correlations and the spin-Peierls
instability in the SU(2) symmetric chain
(i.e., the antiferromagnetic Heisenberg model in zero field),
in which the marginally-irrelevant operator in the low-energy effective
Hamiltonian yields logarithmic corrections.
We have derived the asymptotic formula for the excitation gap in the SU(2)
symmetric chain with bond alternation and numerically determined
the coefficients of the first few terms in the formula expanded in powers of
the coupling constant.
From the formula of the gap, we have obtained the asymptotic power-law
behavior of the dimer correlation function with a multiplicative logarithmic
correction, Eq.\ (\ref{eq:Odim-cor_SU2}).

The dimer correlation amplitudes obtained
in this work can be used for quantitative study of physical properties
related to the dimer operators, such as the spin-Peierls instability,
dynamical structure factors of dimer operators measured in resonant
inelastic x-ray scattering experiments,
the effects of weak interchain dimer-dimer interactions in quasi-1D systems, etc.

\acknowledgments
We thank Temo Vekua, Masahiro Sato, and Tatsuya Nagao for fruitful discussions.
S.L. would like to thank Gennady Y. Chitov for the collaboration on the earlier stage of this work.
T.H. was supported by JSPS KAKENHI Grant Number 15K05198.
The research of S.L. is supported by the NSF under grant number NSF-PHY-1404056.


\begin{thebibliography}{99}
\bibitem{systematic_error}
We note that the leading irrelevant operator $g \cos(\sqrt{8\pi}\phi)$
is not included in the effective Hamiltonian (\ref{eq:Gaussian}) used
in our analysis for $|\Delta| < 1$.
This may induce systematic errors for $\Delta$ close
to unity, in addition to the estimated numerical errors shown
in the Table I.
We expect that the extrapolation of $c_1^a(L)$ to $L\to\infty$
performed in our analysis should lessen potential systematic errors.
\bibitem{diff_gap_d05}
The relative difference between the DMRG data (circles) and
the analytic results (dotted line) in Fig.\ \ref{fig:gap_d05}
is about 1.2\% at $\delta=2^{-10}$, 1.7\% at $\delta=2^{-3}$,
and smaller for intermediate values of $\delta$.
The difference at $\delta=2^{-10}$ is on the same order
as the numerical error in the DMRG data of $E_{\rm g}(\delta)$.
\end{thebibliography}
\end{document}


\title{
Supplemental Material for 
``Dimer correlation amplitudes and dimer excitation gap in spin-1/2 XXZ
and Heisenberg chains"
}
\author{Toshiya Hikihara}
\affiliation{Faculty of Science and Technology, Gunma University, 
Kiryu, Gunma 376-8515, Japan}
\author{Akira Furusaki}
\affiliation{Condensed Matter Theory Laboratory, RIKEN,
Wako, Saitama 351-0198, Japan}
\affiliation{RIKEN Center for Emergent Matter Science 
(CEMS), Wako, Saitama, 351-0198, Japan}
\author{Sergei Lukyanov}
\affiliation{NHETC, Department of Physics and Astronomy, 
Rutgers University, Piscataway, NJ 08855-0849, USA}
\date{\today}

\begin{abstract}
\end{abstract}

\maketitle

In this supplemental material, we present the numerical data of the amplitudes $B_1^\pm$ and $B_1^z$ of the leading oscillating term of the dimer correlation functions.
Tables \ref{tab:B1pm1} - \ref{tab:B1zz3} show the values of $B_1^\pm$ and $B_1^z$ as functions of the anisotropy parameter $\Delta$ and the magnetization $M$.
The data for $-0.6 \le \Delta \le 0.9$ and $M=0$ are the same as those presented in Table\ I of the main text, and some of the data for $M > 0$ are shown in Fig.\ 8 of the main text.
The number in the parentheses represents the error on the last quoted digit(s).

\begin{table}
\caption{
Amplitude $B_1^\pm$ for $-0.6 \le \Delta \le 0.0$.
}
\label{tab:B1pm1}
\begin{center}
\begin{tabular}{cccccccc}
\hline
\hline
 $M$  & $\Delta =-0.6$ & $\Delta =-0.5$ & $\Delta =-0.4$ & $\Delta =-0.3$ & $\Delta =-0.2$ & $\Delta =-0.1$ & $\Delta = 0.0$ \\
\hline
 0.00 & 0.00760(7) & 0.00872(1) & 0.00975(1) & 0.01068(1) & 0.01149(1) & 0.01216(1) & 0.01267(1) \\
\hline
 0.02 &   0.0076    &   0.0087    &   0.0097    &   0.0107    &   0.0115    &   0.0121    &   0.0127    \\
 0.04 &   0.0076    &   0.0087    &   0.0098    &   0.0107    &   0.0115    &   0.0121    &   0.0127    \\
 0.06 &   0.0077    &   0.0088    &   0.0098    &   0.0107    &   0.0114    &   0.0121    &   0.0127    \\
 0.08 &   0.0078    &   0.0088    &   0.0098    &   0.0107    &   0.0114    &   0.0121    &   0.0127    \\
 0.10 &   0.0079    &   0.0089    &   0.0098    &   0.0107    &   0.0114    &   0.0121    &   0.0127    \\
 0.12 &   0.0080    &   0.0090    &   0.0099    &   0.0107    &   0.0114    &   0.0121    &   0.0127    \\
 0.14 &   0.0082    &   0.0091    &   0.0100    &   0.0108    &   0.0115    &   0.0121    &   0.0127    \\
 0.16 &   0.0084    &   0.0093    &   0.0101    &   0.0108    &   0.0115    &   0.0121    &   0.0127    \\
 0.18 &   0.0086    &   0.0095    &   0.0102    &   0.0109    &   0.0116    &   0.0122    &   0.0127    \\
 0.20 &   0.0089    &   0.0097    &   0.0104    &   0.0111    &   0.0117    &   0.0122    &   0.0127    \\
 0.22 &   0.0093    &   0.0100(1) &   0.0106(1) &   0.0112    &   0.0118    &   0.0122    &   0.0127    \\
 0.24 &   0.0097(1) &   0.0103(1) &   0.0109(1) &   0.0114    &   0.0119    &   0.0123    &   0.0127    \\
 0.26 &   0.0102(1) &   0.0107(1) &   0.0112(1) &   0.0116(1) &   0.0120    &   0.0124    &   0.0127    \\
 0.28 &   0.0108(1) &   0.0111(1) &   0.0115(1) &   0.0119(1) &   0.0122    &   0.0124    &   0.0127    \\
 0.30 &   0.0114(2) &   0.0117(1) &   0.0119(1) &   0.0121(1) &   0.0123(1) &   0.0125    &   0.0127    \\
 0.32 &   0.0122(2) &   0.0122(2) &   0.0123(1) &   0.0124(1) &   0.0125(1) &   0.0126    &   0.0127    \\
 0.34 &   0.0131(2) &   0.0129(2) &   0.0128(1) &   0.0127(1) &   0.0127(1) &   0.0127    &   0.0127    \\
 0.36 &   0.0140(3) &   0.0135(2) &   0.0132(1) &   0.0130(1) &   0.0129(1) &   0.0128    &   0.0127    \\
 0.38 &   0.0150(3) &   0.0142(2) &   0.0137(2) &   0.0133(1) &   0.0130(1) &   0.0128    &   0.0127    \\
 0.40 &   0.0159(4) &   0.0148(3) &   0.0141(2) &   0.0135(1) &   0.0132(1) &   0.0129    &   0.0127    \\
 0.42 &   0.0168(4) &   0.0153(3) &   0.0144(2) &   0.0137(1) &   0.0133(1) &   0.0129    &   0.0127    \\
 0.44 &   0.0172(4) &   0.0156(3) &   0.0145(2) &   0.0138(1) &   0.0133(1) &   0.0130    &   0.0127    \\
 0.46 &   0.0170(4) &   0.0154(2) &   0.0144(1) &   0.0138(1) &   0.0133(1) &   0.0129    &   0.0127    \\
 0.48 &   0.0158(3) &   0.0147(2) &   0.0140(1) &   0.0135(1) &   0.0131    &   0.0129    &   0.0127    \\
\hline
\hline
\end{tabular}
\end{center}
\end{table}

\begin{table}
\caption{
Amplitude $B_1^\pm$ for $0.1 \le \Delta \le 0.7$.
}
\label{tab:B1pm2}
\begin{center}
\begin{tabular}{cccccccc}
\hline
\hline
 $M$  & $\Delta =$ 0.1 & $\Delta =$ 0.2 & $\Delta =$ 0.3 & $\Delta =$ 0.4 & $\Delta =$ 0.5 & $\Delta =$ 0.6 & $\Delta =$ 0.7 \\
\hline
 0.00 & 0.01302(1) & 0.01319(1) & 0.01314(1) & 0.01285(1) & 0.01229(2) & 0.01139(2) & 0.01008(2) \\
\hline
 0.02 &   0.0130    &   0.0133    &   0.0133    &   0.0132    &   0.0129(1) &   0.0125(1) &   0.0120(1) \\
 0.04 &   0.0131    &   0.0133    &   0.0135    &   0.0135(1) &   0.0134(1) &   0.0133(1) &   0.0130(2) \\
 0.06 &   0.0131    &   0.0134    &   0.0137    &   0.0138(1) &   0.0139(1) &   0.0139(1) &   0.0138(1) \\
 0.08 &   0.0131    &   0.0135    &   0.0138    &   0.0140(1) &   0.0142(1) &   0.0143(1) &   0.0144(1) \\
 0.10 &   0.0132    &   0.0136    &   0.0139    &   0.0142(1) &   0.0144(1) &   0.0146(1) &   0.0148(1) \\
 0.12 &   0.0132    &   0.0136    &   0.0140    &   0.0143(1) &   0.0146(1) &   0.0148(1) &   0.0150(1) \\
 0.14 &   0.0132    &   0.0136    &   0.0140    &   0.0143(1) &   0.0146(1) &   0.0149(1) &   0.0151(1) \\
 0.16 &   0.0132    &   0.0136    &   0.0140    &   0.0143(1) &   0.0146(1) &   0.0149(1) &   0.0151(1) \\
 0.18 &   0.0131    &   0.0135    &   0.0139    &   0.0142(1) &   0.0145(1) &   0.0148(1) &   0.0150(1) \\
 0.20 &   0.0131    &   0.0135    &   0.0138(1) &   0.0141(1) &   0.0144(1) &   0.0146(1) &   0.0148(1) \\
 0.22 &   0.0130    &   0.0134    &   0.0137(1) &   0.0140(1) &   0.0142(1) &   0.0144(1) &   0.0146(1) \\
 0.24 &   0.0130    &   0.0133    &   0.0135(1) &   0.0138(1) &   0.0140(1) &   0.0142(1) &   0.0143(1) \\
 0.26 &   0.0129    &   0.0132    &   0.0134(1) &   0.0136(1) &   0.0137(1) &   0.0139(1) &   0.0140(1) \\
 0.28 &   0.0129    &   0.0130    &   0.0132(1) &   0.0133(1) &   0.0135(1) &   0.0136(1) &   0.0137(1) \\
 0.30 &   0.0128    &   0.0129    &   0.0130(1) &   0.0131(1) &   0.0132(1) &   0.0133(1) &   0.0133(1) \\
 0.32 &   0.0127    &   0.0128    &   0.0128(1) &   0.0129(1) &   0.0129(1) &   0.0130(1) &   0.0130(1) \\
 0.34 &   0.0127    &   0.0127    &   0.0127(1) &   0.0127(1) &   0.0127(1) &   0.0127(1) &   0.0127(1) \\
 0.36 &   0.0126    &   0.0125    &   0.0125(1) &   0.0125(1) &   0.0124(1) &   0.0124(1) &   0.0124(1) \\
 0.38 &   0.0125    &   0.0124    &   0.0123(1) &   0.0123(1) &   0.0122(1) &   0.0121(1) &   0.0121(1) \\
 0.40 &   0.0125    &   0.0123    &   0.0122(1) &   0.0121(1) &   0.0120(1) &   0.0119(1) &   0.0119(1) \\
 0.42 &   0.0124    &   0.0123    &   0.0121(1) &   0.0120(1) &   0.0119(1) &   0.0118(1) &   0.0117(1) \\
 0.44 &   0.0124    &   0.0122    &   0.0121(1) &   0.0119(1) &   0.0118(1) &   0.0117(1) &   0.0116(1) \\
 0.46 &   0.0124    &   0.0123    &   0.0121    &   0.0120(1) &   0.0119(1) &   0.0118(1) &   0.0117(1) \\
 0.48 &   0.0125    &   0.0124    &   0.0122    &   0.0121    &   0.0121(1) &   0.0120(1) &   0.0119(1) \\
\hline
\hline
\end{tabular}
\end{center}
\end{table}

\begin{table}
\caption{
Amplitude $B_1^\pm$ for $0.8 \le \Delta \le 5.0$.
}
\label{tab:B1pm3}
\begin{center}
\begin{tabular}{cccccccc}
\hline
\hline
 $M$  & $\Delta =$ 0.8 & $\Delta =$ 0.9 & $\Delta =$ 1.0 & $\Delta =$ 1.5 & $\Delta =$ 2.0 & $\Delta =$ 3.0 & $\Delta =$ 5.0 \\
\hline
 0.00 & 0.00826 & 0.00582(6) & - & - & - & - & - \\
\hline
 0.02 &   0.0113(1) &   0.0106(2) &   0.0097(2) &   0.0059(2) &   0.0039(1) &   0.0027(1) &   0.0022    \\
 0.04 &   0.0126(2) &   0.0123(2) &   0.0118(3) &   0.0096(4) &   0.0081(3) &   0.0066(2) &   0.0057(2) \\
 0.06 &   0.0137(2) &   0.0135(2) &   0.0133(2) &   0.0122(3) &   0.0113(3) &   0.0102(3) &   0.0093(2) \\
 0.08 &   0.0144(1) &   0.0144(2) &   0.0144(2) &   0.0140(2) &   0.0136(3) &   0.0130(3) &   0.0124(2) \\
 0.10 &   0.0149(1) &   0.0150(1) &   0.0151(2) &   0.0152(2) &   0.0151(3) &   0.0150(3) &   0.0147(2) \\
 0.12 &   0.0152(1) &   0.0154(1) &   0.0155(2) &   0.0159(2) &   0.0161(3) &   0.0162(3) &   0.0162(3) \\
 0.14 &   0.0153(1) &   0.0155(1) &   0.0157(2) &   0.0163(2) &   0.0166(3) &   0.0170(3) &   0.0172(3) \\
 0.16 &   0.0153(1) &   0.0155(1) &   0.0157(2) &   0.0164(2) &   0.0168(3) &   0.0173(3) &   0.0176(3) \\
 0.18 &   0.0153(1) &   0.0154(1) &   0.0156(1) &   0.0163(2) &   0.0167(3) &   0.0172(3) &   0.0177(3) \\
 0.20 &   0.0150(1) &   0.0152(1) &   0.0154(2) &   0.0160(2) &   0.0165(2) &   0.0170(3) &   0.0174(3) \\
 0.22 &   0.0148(1) &   0.0149(1) &   0.0151(2) &   0.0157(2) &   0.0160(2) &   0.0165(3) &   0.0170(3) \\
 0.24 &   0.0145(1) &   0.0146(1) &   0.0147(2) &   0.0152(2) &   0.0155(2) &   0.0159(3) &   0.0163(3) \\
 0.26 &   0.0141(1) &   0.0142(1) &   0.0143(2) &   0.0147(2) &   0.0150(2) &   0.0153(3) &   0.0156(3) \\
 0.28 &   0.0138(1) &   0.0138(1) &   0.0139(2) &   0.0142(2) &   0.0144(2) &   0.0146(2) &   0.0149(3) \\
 0.30 &   0.0134(1) &   0.0134(1) &   0.0135(1) &   0.0137(2) &   0.0138(2) &   0.0140(2) &   0.0141(3) \\
 0.32 &   0.0130(1) &   0.0131(1) &   0.0131(1) &   0.0132(2) &   0.0132(2) &   0.0133(2) &   0.0134(2) \\
 0.34 &   0.0127(1) &   0.0127(1) &   0.0127(1) &   0.0127(2) &   0.0127(2) &   0.0127(2) &   0.0127(2) \\
 0.36 &   0.0123(1) &   0.0123(1) &   0.0123(1) &   0.0122(2) &   0.0122(2) &   0.0122(2) &   0.0121(2) \\
 0.38 &   0.0121(1) &   0.0120(1) &   0.0120(1) &   0.0119(2) &   0.0118(2) &   0.0117(2) &   0.0116(2) \\
 0.40 &   0.0118(1) &   0.0118(1) &   0.0117(1) &   0.0115(1) &   0.0114(2) &   0.0113(2) &   0.0112(2) \\
 0.42 &   0.0116(1) &   0.0116(1) &   0.0115(1) &   0.0113(1) &   0.0112(2) &   0.0110(2) &   0.0108(2) \\
 0.44 &   0.0116(1) &   0.0115(1) &   0.0114(1) &   0.0112(1) &   0.0111(1) &   0.0109(2) &   0.0107(2) \\
 0.46 &   0.0116(1) &   0.0115(1) &   0.0115(1) &   0.0113(1) &   0.0111(1) &   0.0109(1) &   0.0108(1) \\
 0.48 &   0.0119(1) &   0.0118(1) &   0.0118(1) &   0.0116(1) &   0.0115(1) &   0.0113(1) &   0.0112(1) \\
\hline
\hline
\end{tabular}
\end{center}
\end{table}

\begin{table}
\caption{
Amplitude $B_1^z$ for $-0.6 \le \Delta \le 0.0$.
}
\label{tab:B1zz1}
\begin{center}
\begin{tabular}{cccccccc}
\hline
\hline
 $M$  & $\Delta =-0.6$ & $\Delta =-0.5$ & $\Delta =-0.4$ & $\Delta =-0.3$ & $\Delta =-0.2$ & $\Delta =-0.1$ & $\Delta = 0.0$ \\
\hline
 0.00 & 0.0190(3) & 0.01999(3) & 0.02066(2) & 0.02103(2) & 0.02112(2) & 0.02095(2) & 0.02055(2) \\
\hline
 0.02 &   0.0190    &   0.0200    &   0.0207    &   0.0211    &   0.0211    &   0.0210    &   0.0206    \\
 0.04 &   0.0191(1) &   0.0201(1) &   0.0208    &   0.0212    &   0.0213    &   0.0212    &   0.0208    \\
 0.06 &   0.0193(1) &   0.0203(1) &   0.0210    &   0.0214    &   0.0215    &   0.0215    &   0.0212    \\
 0.08 &   0.0197(1) &   0.0206(1) &   0.0213    &   0.0217    &   0.0219    &   0.0219    &   0.0218    \\
 0.10 &   0.0201(1) &   0.0210    &   0.0217    &   0.0222    &   0.0224    &   0.0225    &   0.0225    \\
 0.12 &   0.0207(1) &   0.0216    &   0.0223    &   0.0228    &   0.0231    &   0.0233    &   0.0234    \\
 0.14 &   0.0214(1) &   0.0222    &   0.0230    &   0.0235    &   0.0239    &   0.0242    &   0.0244    \\
 0.16 &   0.0222(1) &   0.0231    &   0.0238    &   0.0244    &   0.0249    &   0.0253    &   0.0256    \\
 0.18 &   0.0233(1) &   0.0242(1) &   0.0249(1) &   0.0255(1) &   0.0261    &   0.0266    &   0.0270    \\
 0.20 &   0.0246(1) &   0.0254(1) &   0.0261(1) &   0.0268(1) &   0.0275(1) &   0.0280    &   0.0285    \\
 0.22 &   0.0262(1) &   0.0269(1) &   0.0277(1) &   0.0284(1) &   0.0290(1) &   0.0296(1) &   0.0301    \\
 0.24 &   0.0280(2) &   0.0287(2) &   0.0294(2) &   0.0301(1) &   0.0307(1) &   0.0313    &   0.0318    \\
 0.26 &   0.0302(2) &   0.0307(2) &   0.0314(2) &   0.0320(1) &   0.0326(1) &   0.0331    &   0.0336    \\
 0.28 &   0.0328(4) &   0.0332(3) &   0.0337(3) &   0.0342(2) &   0.0347(1) &   0.0351(1) &   0.0355    \\
 0.30 &   0.0361(6) &   0.0361(5) &   0.0363(4) &   0.0366(3) &   0.0369(2) &   0.0372(1) &   0.0374    \\
 0.32 &   0.0398(6) &   0.0393(5) &   0.0392(4) &   0.0392(3) &   0.0392(2) &   0.0393(1) &   0.0393    \\
 0.34 &   0.0440(7) &   0.0429(6) &   0.0422(4) &   0.0418(3) &   0.0416(2) &   0.0414(1) &   0.0412    \\
 0.36 &   0.0487(9) &   0.0467(7) &   0.0454(5) &   0.0446(3) &   0.0439(2) &   0.0434(1) &   0.0431    \\
 0.38 &   0.0538(12) &   0.0507(8) &   0.0487(6) &   0.0473(4) &   0.0462(2) &   0.0454(1) &   0.0448    \\
 0.40 &   0.0589(14) &   0.0545(10) &   0.0517(7) &   0.0498(4) &   0.0483(3) &   0.0473(1) &   0.0464    \\
 0.42 &   0.0637(16) &   0.0580(11) &   0.0544(7) &   0.0520(5) &   0.0502(3) &   0.0489(1) &   0.0479(1) \\
 0.44 &   0.0667(15) &   0.0602(10) &   0.0562(6) &   0.0535(4) &   0.0516(2) &   0.0501(1) &   0.0490    \\
 0.46 &   0.0671(14) &   0.0608(9) &   0.0569(6) &   0.0543(3) &   0.0524(2) &   0.0510(1) &   0.0499    \\
 0.48 &   0.0631(12) &   0.0585(8) &   0.0557(5) &   0.0538(3) &   0.0524(2) &   0.0513(1) &   0.0505    \\
\hline
\hline
\end{tabular}
\end{center}
\end{table}

\begin{table}
\caption{
Amplitude $B_1^z$ for $0.1 \le \Delta \le 0.7$.
}
\label{tab:B1zz2}
\begin{center}
\begin{tabular}{cccccccc}
\hline
\hline
 $M$  & $\Delta =$ 0.1 & $\Delta =$ 0.2 & $\Delta =$ 0.3 & $\Delta =$ 0.4 & $\Delta =$ 0.5 & $\Delta =$ 0.6 & $\Delta =$ 0.7 \\
\hline
 0.00 & 0.01992(2) & 0.01909(1) & 0.01804(1) & 0.01677(2) & 0.01528(3) & 0.01351(3) & 0.01144(2) \\
\hline
 0.02 &   0.0200    &   0.0193    &   0.0184    &   0.0174(1) &   0.0163(1) &   0.0151(1) &   0.0138(1) \\
 0.04 &   0.0204    &   0.0197(1) &   0.0190(1) &   0.0182(1) &   0.0173(2) &   0.0164(2) &   0.0155(2) \\
 0.06 &   0.0209(1) &   0.0204(1) &   0.0199(1) &   0.0192(1) &   0.0186(1) &   0.0179(2) &   0.0173(2) \\
 0.08 &   0.0216    &   0.0213(1) &   0.0209(1) &   0.0205(1) &   0.0201(1) &   0.0196(2) &   0.0192(2) \\
 0.10 &   0.0224    &   0.0223(1) &   0.0221(1) &   0.0219(1) &   0.0216(1) &   0.0214(1) &   0.0211(2) \\
 0.12 &   0.0235    &   0.0235(1) &   0.0234(1) &   0.0233(1) &   0.0233(1) &   0.0232(2) &   0.0230(2) \\
 0.14 &   0.0246(1) &   0.0247(1) &   0.0248(1) &   0.0249(1) &   0.0249(1) &   0.0250(2) &   0.0250(2) \\
 0.16 &   0.0259(1) &   0.0261(1) &   0.0263(1) &   0.0265(2) &   0.0267(2) &   0.0268(2) &   0.0269(2) \\
 0.18 &   0.0274    &   0.0277(1) &   0.0280(1) &   0.0283(1) &   0.0285(1) &   0.0287(2) &   0.0289(2) \\
 0.20 &   0.0289    &   0.0293(1) &   0.0297(1) &   0.0300(1) &   0.0302(2) &   0.0305(2) &   0.0307(2) \\
 0.22 &   0.0306    &   0.0310(1) &   0.0313(1) &   0.0317(1) &   0.0320(2) &   0.0323(2) &   0.0325(2) \\
 0.24 &   0.0323    &   0.0327(1) &   0.0330(1) &   0.0334(2) &   0.0337(2) &   0.0339(2) &   0.0342(3) \\
 0.26 &   0.0340(1) &   0.0344(1) &   0.0347(1) &   0.0350(2) &   0.0353(2) &   0.0355(3) &   0.0357(3) \\
 0.28 &   0.0358    &   0.0361(1) &   0.0364(1) &   0.0366(2) &   0.0368(2) &   0.0370(3) &   0.0372(3) \\
 0.30 &   0.0376    &   0.0378(1) &   0.0380(1) &   0.0382(2) &   0.0383(2) &   0.0384(3) &   0.0386(3) \\
 0.32 &   0.0394    &   0.0395(1) &   0.0395(1) &   0.0396(2) &   0.0397(2) &   0.0397(3) &   0.0398(3) \\
 0.34 &   0.0411(1) &   0.0410(1) &   0.0410(2) &   0.0409(2) &   0.0409(3) &   0.0408(3) &   0.0408(4) \\
 0.36 &   0.0428(1) &   0.0425(1) &   0.0423(2) &   0.0422(2) &   0.0420(3) &   0.0419(3) &   0.0418(3) \\
 0.38 &   0.0443(1) &   0.0439(1) &   0.0436(2) &   0.0433(2) &   0.0430(3) &   0.0428(3) &   0.0426(3) \\
 0.40 &   0.0457(1) &   0.0452(1) &   0.0447(2) &   0.0443(2) &   0.0440(3) &   0.0437(3) &   0.0434(3) \\
 0.42 &   0.0470    &   0.0463(1) &   0.0458(2) &   0.0453(2) &   0.0449(3) &   0.0445(3) &   0.0442(3) \\
 0.44 &   0.0481(1) &   0.0473(1) &   0.0467(2) &   0.0462(2) &   0.0457(3) &   0.0453(3) &   0.0450(3) \\
 0.46 &   0.0490(1) &   0.0483(1) &   0.0477(2) &   0.0472(2) &   0.0467(3) &   0.0463(3) &   0.0460(3) \\
 0.48 &   0.0498    &   0.0492(1) &   0.0488(1) &   0.0484(2) &   0.0480(2) &   0.0478(2) &   0.0475(2) \\
\hline
\hline
\end{tabular}
\end{center}
\end{table}

\begin{table}
\caption{
Amplitude $B_1^z$ for $0.8 \le \Delta \le 5.0$.
}
\label{tab:B1zz3}
\begin{center}
\begin{tabular}{cccccccc}
\hline
\hline
 $M$  & $\Delta =$ 0.8 & $\Delta =$ 0.9 & $\Delta =$ 1.0 & $\Delta =$ 1.5 & $\Delta =$ 2.0 & $\Delta =$ 3.0 & $\Delta =$ 5.0 \\
\hline
 0.00 & 0.00898 & 0.00606(7) & - & - & - & - & - \\
\hline
 0.02 &   0.0125(2) &   0.0112(2) &   0.0100(2) &   0.0052(2) &   0.0033(1) &   0.0023    &   0.0019    \\
 0.04 &   0.0145(3) &   0.0136(3) &   0.0128(3) &   0.0093(4) &   0.0074(3) &   0.0060(2) &   0.0053(1) \\
 0.06 &   0.0166(2) &   0.0160(2) &   0.0154(3) &   0.0130(3) &   0.0115(3) &   0.0102(2) &   0.0094(2) \\
 0.08 &   0.0187(2) &   0.0183(2) &   0.0179(2) &   0.0163(3) &   0.0153(3) &   0.0143(2) &   0.0137(2) \\
 0.10 &   0.0209(2) &   0.0206(2) &   0.0204(2) &   0.0194(3) &   0.0188(3) &   0.0182(3) &   0.0179(3) \\
 0.12 &   0.0229(2) &   0.0228(2) &   0.0227(2) &   0.0224(3) &   0.0221(3) &   0.0219(3) &   0.0218(3) \\
 0.14 &   0.0250(2) &   0.0250(2) &   0.0250(3) &   0.0251(3) &   0.0251(4) &   0.0253(4) &   0.0255(4) \\
 0.16 &   0.0270(3) &   0.0271(3) &   0.0272(3) &   0.0276(4) &   0.0278(4) &   0.0282(5) &   0.0286(5) \\
 0.18 &   0.0291(2) &   0.0292(2) &   0.0294(3) &   0.0299(3) &   0.0304(4) &   0.0309(5) &   0.0315(5) \\
 0.20 &   0.0310(2) &   0.0312(3) &   0.0313(3) &   0.0321(4) &   0.0326(4) &   0.0333(5) &   0.0339(5) \\
 0.22 &   0.0328(2) &   0.0330(3) &   0.0332(3) &   0.0340(4) &   0.0345(4) &   0.0352(5) &   0.0359(5) \\
 0.24 &   0.0344(3) &   0.0346(3) &   0.0348(3) &   0.0356(4) &   0.0361(5) &   0.0368(5) &   0.0375(6) \\
 0.26 &   0.0360(3) &   0.0361(3) &   0.0363(4) &   0.0370(5) &   0.0374(5) &   0.0380(6) &   0.0387(6) \\
 0.28 &   0.0374(3) &   0.0375(3) &   0.0377(4) &   0.0382(5) &   0.0386(5) &   0.0390(6) &   0.0395(7) \\
 0.30 &   0.0387(3) &   0.0388(3) &   0.0389(4) &   0.0392(5) &   0.0395(5) &   0.0398(6) &   0.0401(7) \\
 0.32 &   0.0398(3) &   0.0399(4) &   0.0399(4) &   0.0401(5) &   0.0402(5) &   0.0403(6) &   0.0405(7) \\
 0.34 &   0.0408(4) &   0.0408(4) &   0.0407(4) &   0.0407(5) &   0.0407(6) &   0.0406(7) &   0.0406(7) \\
 0.36 &   0.0417(4) &   0.0416(4) &   0.0415(4) &   0.0412(5) &   0.0411(6) &   0.0409(6) &   0.0406(7) \\
 0.38 &   0.0425(4) &   0.0423(4) &   0.0422(4) &   0.0417(5) &   0.0414(6) &   0.0410(6) &   0.0407(7) \\
 0.40 &   0.0432(4) &   0.0430(4) &   0.0428(4) &   0.0422(5) &   0.0417(6) &   0.0412(6) &   0.0407(7) \\
 0.42 &   0.0440(4) &   0.0437(4) &   0.0435(4) &   0.0427(5) &   0.0422(5) &   0.0415(6) &   0.0409(7) \\
 0.44 &   0.0447(4) &   0.0444(4) &   0.0442(4) &   0.0433(5) &   0.0427(5) &   0.0420(6) &   0.0413(6) \\
 0.46 &   0.0457(3) &   0.0455(4) &   0.0452(4) &   0.0444(4) &   0.0438(5) &   0.0431(5) &   0.0424(6) \\
 0.48 &   0.0473(3) &   0.0471(3) &   0.0469(3) &   0.0462(4) &   0.0458(4) &   0.0452(4) &   0.0447(5) \\
\hline
\hline
\end{tabular}
\end{center}
\end{table}